\documentclass{article}

\usepackage{arxiv}

\usepackage[utf8]{inputenc}
\usepackage[T1]{fontenc}
\usepackage{amsmath,amssymb,amsthm,mathtools}
\usepackage{amsfonts}
\usepackage{booktabs,array,multirow}
\usepackage{graphicx}
\graphicspath{{figures/}}
\usepackage{subcaption}
\usepackage{nicefrac}
\usepackage{microtype}
\usepackage{enumitem}
\usepackage{xcolor}
\usepackage{float}
\usepackage{placeins}
\usepackage{bm}
\usepackage{bbm}
\usepackage[hidelinks]{hyperref}
\usepackage{url}
\usepackage[nameinlink,noabbrev]{cleveref}
\usepackage[sort&compress]{natbib}
\usepackage{doi}
\usepackage{authblk}

\newtheorem{theorem}{Theorem}
\newtheorem{proposition}[theorem]{Proposition}

\newtheorem{assumption}{Assumption}
\theoremstyle{definition}

\theoremstyle{remark}

\newcommand{\E}{\mathbb{E}}
\newcommand{\R}{\mathbb{R}}
\newcommand{\Ptwo}{\mathcal{P}_2}
\newcommand{\dd}{\,\mathrm{d}}

\newcommand{\1}{\mathbbm{1}}

\newcommand{\eps}{\varepsilon}
\newcommand{\Tspace}{\mathcal T}
\newcommand{\Bop}{\mathcal B}
\newcommand{\Lop}{\mathcal L}

\title{A Dynamic Model of Party Differentiation under Electoral Competition}
\author[1]{Junjie Liu}
\author[2, 3]{Wencheng Lin}

\affil[1]{Department of Political Science, Trinity College Dublin, Dublin, Ireland}
\affil[2]{Guangdong Provincial Key Laboratory of Interdisciplinary Research and Application for Data Science, Zhuhai, China}
\affil[3]{Beijing Normal--Hong Kong Baptist University, Zhuhai, China}
\affil[ ]{\texttt{liuj13@tcd.ie, wenchenlin@bnbu.edu.cn}}

\date{September, 2026}

\renewcommand{\shorttitle}{}
\renewcommand{\headeright}{}
\renewcommand{\undertitle}{}

\hypersetup{
  pdftitle={A Dynamic Model of Party Differentiation under Electoral Competition},
  pdfauthor={Junjie Liu, Wencheng Lin},
  pdfkeywords={party differentiation, electoral competition,endogenous turnout, electoral geography, Wasserstein dynamics, symmetry breaking}
}

\begin{document}
\maketitle

\begin{abstract}
Why do political parties polarise when most voters remain in the centre? Moving beyond conventional point-based models, we address this classic puzzle by developing a spatial voting framework that conceptualises political parties as continuous ideological distributions. Operating under endogenous turnout, these parties adapt their positioning to maximise a combination of total votes and parliamentary representation. We demonstrate that this theoretical formulation reveals a distinct phase transition in multiparty competition, moving from centripetal convergence to centrifugal separation. These dual objectives do not necessarily reward the same directional party movement as voters only cast ballots for ideologically acceptable parties. The numerical results map the precise boundaries of this transition: a higher party count destabilises the centrist equilibrium and expands the zone of divergence, whereas pronounced internal party dispersion and a stronger emphasis on representation contract it. Furthermore, electoral institutions critically govern this divergence: when voter preferences vary across geographic districts, the same national electorate can trigger entirely different phase transitions depending on how votes are converted into seats. Together, our results show how party differentiation emerges endogenously from the dynamics of electoral competition itself. Whether small initial differences grow or decay is not merely a reflection of voter preferences, but an outcome dictated by party goals, competitor count, internal party breadth, and vote-to-seat conversion rules.\footnote{The source code is openly available in the \href{https://github.com/TerenceLiu98/dynamic-model-multiparty-competition}{GitHub Repository}.}
\end{abstract}


\section{Introduction}
\label{sec:introduction}

Parties competing for the same electorate can offer overlapping platforms or occupy distinct ideological positions. They also contain candidates, elites and factions with different views, so the distance between parties and the breadth within each party describe separate dimensions of political competition~\citep{KollnPolk2024,Imre2023,Catalinac2018,MatakosEtAl2024}. A common political centre may therefore support several broad, overlapping parties, while a differentiated system may combine separated centres with substantial internal variation. The question is what makes competition sustain one configuration or amplify differences towards another. The puzzle is particularly clear when voters are symmetrically concentrated around the centre. Moving towards that centre provides access to a large electorate, but also increases overlap with competing parties. Even a small departure from the centre reshapes both the segment of voters a party can reach and the competitors with which it shares those voters. In canonical spatial models, these trade-offs generate incentives for parties to converge toward electorally attractive positions, although the strength and stability of convergence depend on the distribution of voters and the structure of multiparty competition~\citep{Hotelling1929,Downs1957,Petersen1991,SchofieldSenedNixon1998,Schofield2007}. These incentives need not operate uniformly across electoral environments. Platform choices interact with electoral rules and participation, which affect both the composition of the electorate and the electoral return to support at different points in the policy space~\citep{Ortuno1997,Matakos2016,OpreaMartinBrennan2024}. Moreover, when several parties compete simultaneously, a change in one party's position alters not only its own electoral appeal but also the strategic opportunities facing its rivals, including incentives for entry and repositioning~\citep{Adams1998,Bol2019}. Spatial models therefore treat party competition as an interdependent and potentially dynamic process in which parties adapt to changes in the surrounding electoral landscape~\citep{KollmanMillerPage1998,Laver2005}. Platform incentives ultimately depend on participation, the number and location of competitors, and the electoral return to the resulting support. In this research, we specifically ask: \emph{When do small differences between party profiles decay, and when do they instead develop into sustained separation?}

Our analytical perspective is to treat differentiation as feedback from marginal electoral returns. Parties move ideological mass towards positions that improve their own electoral payoff. Voters' choices determine support and participation, electoral institutions convert those votes into parliamentary representation, and the party payoff specifies how parties evaluate these outcomes.. This connection between microscopic choice and collective organisation places the problem within statistical-physics accounts of consensus, clustering and opinion dynamics~\citep{CastellanoFortunatoLoreto2009,HegselmannKrause2002,AguilarJanitaEtAl2024}. Non-local drift and diffusion provide a distributional description of the competition between directed adjustment and dispersion~\citep{Toscani2006,PinedaToralHernandezGarcia2009,CahillGottwald2025}.

The electoral value of support depends on how votes translate into representation. Electoral coordination concerns the concentration of support needed to secure representation under a given rule~\citep{Cox1997}. We examine the platform incentives generated by attracting votes and by their institutional returns. The same increase in votes can have different seat payoffs depending on competing votes, district composition and the allocation rule. In homogeneous proportional representation (PR), a party with unconditional support $q_i$ receives $S_i=q_i/T$, where $T$ is turnout. Let $D_i$ denote a first-order unilateral variation of party $i$'s profile, with all competing profiles held fixed. The quotient rule then gives
\begin{equation}
 D_iS_i=\frac{D_iq_i-S_iD_iT}{T}.
 \label{eq:objective-angle}
\end{equation}
The denominator is itself an outcome of competition. Attracting abstainers can enlarge the electorate that participates, while attracting a competitor's voters redistributes support within it. Even in this proportional benchmark, endogenous turnout makes the reward for acquiring votes differ from its normalised seat return. We make the relative weight of these two rewards explicit in the party objective and ask how it changes the incentives associated with party number.

The model represents each party as an ideological distribution. Satisficing choice links voter acceptance to support and abstention~\citep{YangEtAl2020}, while Wasserstein adaptation expresses movement of entire party profiles towards higher electoral returns~\citep{LanzettiHajarDorfler2022}. An exact choice formula accommodates any fixed number of parties. Entropy regularisation generates internal diffusion, allowing a common stationary profile to have positive width. Electoral rules and district densities enter the same feedback through the conversion of votes into seats. This formulation therefore provides a common framework for asking when a shared political centre is stable, how that stability depends on the number of parties and institutional environment, and how instability develops once differentiation becomes possible. We analyse these questions in two connected parts. Under homogeneous PR, party number and the institutional-payoff weight jointly determine the feedback acting on reflection-odd differences between parties. Numerical phase diagrams then locate its competition with diffusion, and full dynamics follow selected instabilities to their endpoints. Under seat-share maximisation, we identify how electoral rules contribute through marginal responsiveness and geographic curvature. Together, these analyses distinguish three aspects that need not coincide: the existence of a differentiation channel, the range in which it destabilises a common profile, and the speed and eventual outcome of that instability. 

The paper is structured as follows. Section~\ref{sec:model} introduces the model, and Section~\ref{sec:distributional} develops the stability mechanism. Sections~\ref{sec:theta-results} and~\ref{sec:institution} present the numerical and institutional results, respectively, while Section~\ref{sec:discussion} discusses their broader implications and concludes.

\section{Electoral feedback and party adaptation}
\label{sec:model}

In this section, we formulate the coupled dynamic system governing party adaptation as a multi-species transport--difussion process on the ideological manifold $\Omega$. The model integrates microscopic voter choice, macroscopic institutional aggregation, and gradient-flow optimisation into a closed feedback loop. First, each political party is described as a continuous probability density evolving under endogenous voter acceptance and satisficing choice. Second, district-level votes are mapped to parliamentary seat shares via institutional allocation rules, generating the macroscopic return. Finally, parties perform Wasserstein gradient ascent on their objective functionals against entropic diffusion, yielding a system of coupled, non-local Fokker--Planck equations. We detail these three constitutive stages below.

\subsection{Party profiles and voter choice}

There are $K\geq2$ parties and $D\geq1$ districts. Voter densities and party number are fixed. On a compact ideological domain $\Omega$, district $d$ has normalised density $\rho_d$, with $\int_\Omega\rho_d(x)\dd x=1$, and weight $w_d>0$, with $\sum_dw_d=1$. We use the same weights for district electorate size and fixed district seat allocations. The numerical districts have equal weights. Party $i$ has state $\mu_i$ in the $i$-th factor $\mathcal{P}_2(\Omega)$. The joint party state is: 
\begin{equation}
 \bm\mu=(\mu_1,\ldots,\mu_i, \ldots, \mu_K)\in[\Ptwo(\Omega)]^K.
 \label{eq:party-state}
\end{equation}
Here $\Ptwo$ denotes probability measures with finite second moment, the space for quadratic Wasserstein transport. The subscript specifies the moment order; the product has $K$ party components. Each measure has unit mass, describing relative ideological composition within a party. With a density $f_i$, its mean records location and its variance records internal breadth. A point platform is $\mu_i=\delta_{y_i}$.

Voters screen parties for acceptability and then select within the accepted set, a consideration-set formulation of electoral choice~\citep{Simon1955,BendorMookherjeeRay2006,BendorEtAl2011,OscarssonRosema2019}. A voter at $x$ accepts party $i$ with probability
\begin{equation}
 a_i(x)=\int_\Omega K_\sigma(x,y)\dd\mu_i(y),
 \qquad K_\sigma(x,y)=\exp\!\left[-\frac{\lVert x-y\rVert^2}{2\sigma^2}\right].
 \label{eq:avg-sat}
\end{equation}
The tolerance $\sigma$ sets how quickly compatibility falls with ideological distance. Conditional on $x$, acceptance events are independent across parties. The voter chooses uniformly among accepted parties and abstains when the set is empty. The resulting exact probabilities are
\begin{equation}
 p_i(a)=a_i\int_0^1\prod_{j\ne i}[1-(1-z)a_j]\dd z,
 \qquad p_0(a)=\prod_j(1-a_j).
 \label{eq:choice}
\end{equation}
Thus $p_0+\sum_ip_i=1$. Party number enters through the alternatives that can share a voter's acceptable set. The integral avoids enumeration of all $2^K$ sets and is exactly evaluated by a $\lceil K/2\rceil$-node Gauss--Legendre rule in exact arithmetic. Appendix~\ref{app:choiceproof} gives the probabilistic derivation.

\subsection{Support, participation and representation}

The choice probabilities determine district support, turnout and valid-vote shares:
\begin{align}
 q_{id}&=\int_\Omega\rho_d(x)p_i(a(x))\dd x,
 &T_d&=\sum_iq_{id}=1-\int_\Omega\rho_d(x)p_0(a(x))\dd x,
 \label{eq:uncondvotes}\\
 v_{id}&=q_{id}/T_d,
 &\bar q_i&=\sum_dw_dq_{id}.
 \label{eq:validshare}
\end{align}
The 1 in turnout is the normalised mass of potential voters in the district: participation equals this total minus abstention. Unconditional support $q_{id}$ is the fraction of potential voters choosing party $i$, whereas $v_{id}$ conditions on casting a valid vote. The aggregate $\bar q_i$ measures support across the whole electorate. Its sum across parties is aggregate turnout, which changes as profiles evolve.

Let $\mathcal S_i^R(\bm v)$ be the party's national seat share under rule $R$. In the district PR benchmark,
\begin{equation}
 \mathcal S_i^{\mathrm{PR}}=\bar v_i,\qquad
 \bar v_i=\sum_dw_dv_{id}.
 \label{eq:pr}
\end{equation}
Each district's fixed seat weight is allocated proportionally to its valid votes. Nationally pooled PR would give $(\sum_dw_dq_{id})/(\sum_dw_dT_d)$, weighting districts by their contribution to valid votes. These coincide in a homogeneous electorate and can differ when turnout varies across districts. The geographic comparisons use the fixed-weight aggregation in Eq.~\eqref{eq:pr}.

For single-member plurality, the smooth probability of winning district $d$ and its aggregate are
\begin{equation}
 c_{id}=\frac{e^{\beta v_{id}}}{\sum_je^{\beta v_{jd}}},\qquad
 C_i=\sum_dw_dc_{id},\qquad \mathcal S_i^{\mathrm{FPTP}}=C_i.
 \label{eq:fptp}
\end{equation}
The parameter $\beta$ controls the sharpness of district competition. Away from ties, $\beta\to\infty$ gives winner-take-all plurality. A stylised mixed-member proportional (MMP) map combines proportional entitlement $P_i$ with a direct-mandate floor $d_i=\alpha C_i$:
\begin{equation}
 u_i=P_i+\gamma^{-1}\log[1+e^{\gamma(d_i-P_i)}],
 \qquad \mathcal S_i^{\mathrm{MMP}}=u_i/\sum_ju_j.
 \label{eq:mmp}
\end{equation}
Without a threshold, $P_i=\mathcal S_i^{\mathrm{PR}}=\bar v_i$. The constituency-tier weight is $\alpha$, and $\gamma$ controls smoothing of the entitlement floor; as $\gamma \to \infty, u_i \to \max\{P_i, d_i\}$. This creates compensatory and overhang-like responses. Homogeneous analytical PR is threshold-free. The geographic numerical PR and MMP maps apply a smooth five-percent threshold to $\bar v$, specified in Appendix~\ref{app:numerics}. These differentiable allocation models permit the local seat responses needed for continuous adaptation.

\subsection{Objectives and ideological evolution}

Parties adapt to the rewards from acquiring votes and from the parliamentary representation those votes generate:

\begin{equation}
J_i^{R,\theta}=(1-\theta)\bar q_i+\theta\mathcal S_i^R+\eps\mathcal{H}(f_i),
 \qquad \mathcal H(f_i)=-\int_\Omega f_i\log f_i\dd y.
 \label{eq:payoff}
\end{equation}

Here, the \emph{institutional-payoff weight} $\theta\in[0,1]$ interpolates between raw vote mobilisation and parliamentary seat returns, capturing the fundamental trade-off between vote-seeking and office-seeking behaviour~\citep{strom1990behavioral}. Setting $\theta = 0$ corresponds to pure vote maximisation, reflecting a mobilisation-driven strategy where parties evaluate platform movements solely by their expected support ($\bar{q}_i$) and seek a popular mandate by activating disaffected non-voters. Conversely, $\theta = 1$ represents pure seat-share maximisation, formalising the logic of a ``cartelised" party system~\citep{katz2009cartel} where elites compete strictly for legislative leverage ($\mathcal S_i^R$) in a zero-sum parliamentary arena under rule $R$. Because a platform shift alters both aggregate turnout and the geographic distribution of support across districts, intermediate values of $\theta$ capture parties that simultaneously value popular mandates and legislative power. This linear combination provides a transparent behavioural continuum shared by all competitors.

Complementing this strategic drive, the entropy term $\mathcal H(f_i)$ describes persistent internal platform dispersion resulting from candidate turnover, decentralised nominations, and factional competition. Its coefficient $\eps>0$ governs internal diffusion, preventing profiles from collapsing into point masses. The strategic component, by contrast, transports ideological mass toward regions of higher marginal payoff via Wasserstein gradient ascent~\citep{JKO1998,AmbrosioGigliSavare2008,Santambrogio2017}. Combining these forces yields the effective strategic potential
\begin{equation}
 \psi_i^{R,\theta}=(1-\theta)\phi_i+\theta\psi_i^R,
 \qquad \phi_i=\frac{\delta\bar q_i}{\delta\mu_i},\qquad
 \psi_i^R=\frac{\delta\mathcal S_i^R}{\delta\mu_i}.
 \label{eq:mixed-potential}
\end{equation}
The potential in Eq.\ref{eq:mixed-potential} gives the strategic part of each party's unilateral first variation. For positive densities, $\frac{\delta \mathcal{H}}{\delta f_i} = -(1 + \log(f_i))$. Wasserstein ascent of the full objective therefore yields the transport--diffusion dynamics: 

\begin{equation}
 \partial_tf_i+\nabla\cdot(f_i\nabla\psi_i^{R,\theta})=\eps\Delta f_i,
 \qquad (f_i\nabla\psi_i^{R,\theta}-\eps\nabla f_i)\cdot n=0.
 \label{eq:pde}
\end{equation}
The boundary condition imposes zero flux. Each profile remains normalised and non-negative. Parties adapt simultaneously and myopically; their potentials depend on every party profile, creating a coupled game. Smooth kernels and seat maps yield a unique weak solution on each finite time interval under the assumptions in Appendix~\ref{app:wellposed}.

\begin{figure}[t]
\centering
\includegraphics[width=\textwidth]{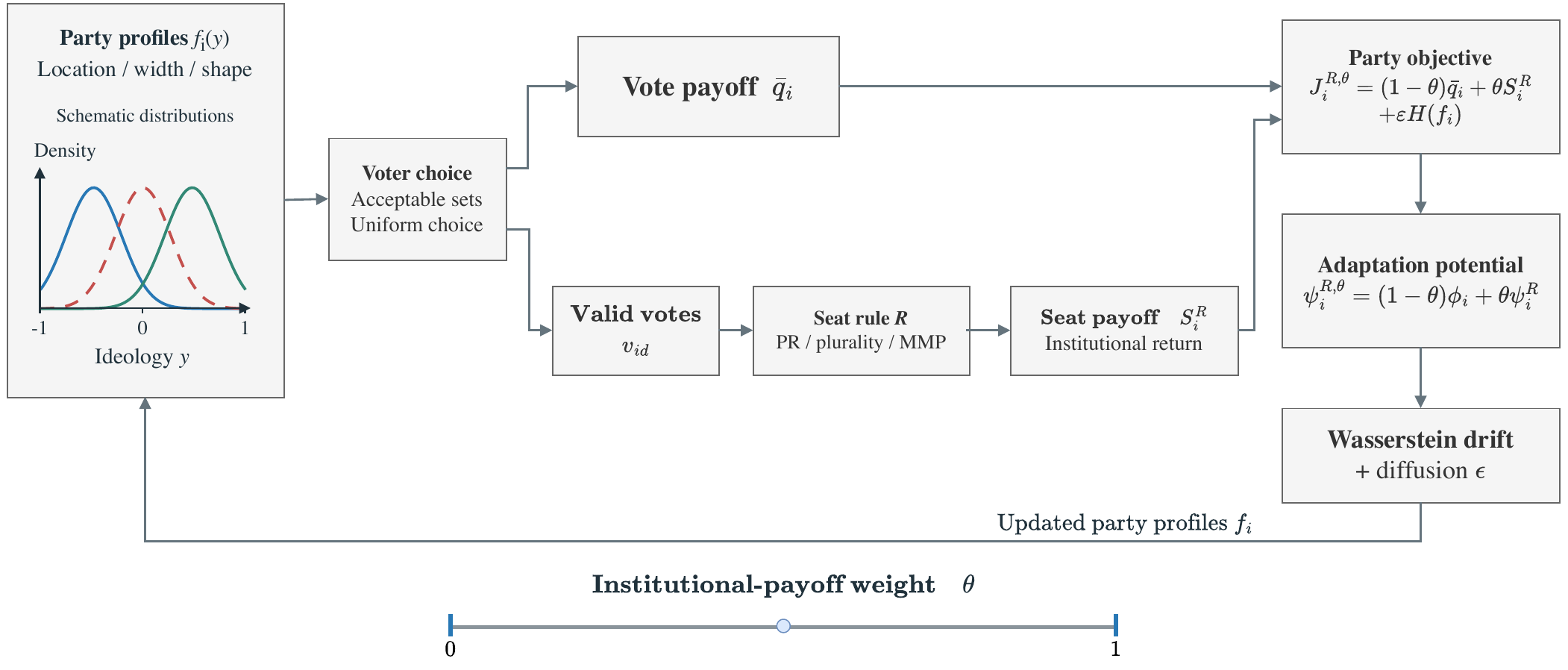}
\caption{\textbf{Institutional-payoff electoral feedback.} Party profiles determine voter acceptance, choice and abstention. The resulting votes enter the objective directly through $\bar q_i$ and through their institutional seat return $S_i^R$. The weight $\theta$ combines these rewards; their marginal potential generates Wasserstein drift, while entropy supplies diffusion. The profile sketches are schematic. PR denotes the district-weighted map defined in the text.}
\label{fig:dynamics-architecture}
\end{figure}

Figure~\ref{fig:dynamics-architecture} summarises this feedback mechanism. For interpretation, the marginal seat response has three factors. A change in acceptance alters voter choice; the turnout correction translates raw support into valid-vote shares; the seat-map derivative translates valid votes into representation. Appendix~\ref{app:firstvariationproof} defines these coefficients, explains their indices and gives the complete chain rule. In homogeneous PR, Eq.~\eqref{eq:objective-angle} is the simplest expression of this mechanism.

For one-dimensional profiles, we track means $m_i=\int yf_i(y)\dd y$, within-party variances $V_i=\int(y-m_i)^2f_i(y)\dd y$, and the seat-weighted between-party spread
\begin{equation}
 P=\left[\sum_i\mathcal S_i^R(m_i-\bar m_S)^2\right]^{1/2},
 \qquad \bar m_S=\sum_i\mathcal S_i^Rm_i.
 \label{eq:observables}
\end{equation}
Profile shapes and overlap complement these moments. A separated pair of means can coexist with broad parties, while equal means can conceal differences in asymmetry or modality. Aggregate turnout $\sum_i\bar q_i$ records how these platform changes alter participation.

\section{When a common political centre becomes unstable}
\label{sec:distributional}\label{subsec:theta-stability}

To investigate the onset of party differentiation, we take as our reference state an unpolarised baseline where all competitors ahre a common ideological profile of positive width. Specifically, consider homogeneous exact PR over an even electorate and an interval $\Omega=[-L,L]$, measuring ideology in voter-standard-deviation units and write $r=\sigma/\sigma_0$. At a diagonal state, all parties share $f$, with acceptance profile $A_f(x)=\int K_\sigma(x,y)f(y)\dd y$ and aggregate turnout $Q_f=\int\rho(x)[1-(1-A_f(x))^K]\dd x$.

Under this symmetry, the party-specific marginal potential in Eq.~\eqref{eq:mixed-potential} collapses to a shared strategic potential:

\begin{align}
 \phi_f(y)
 &=\int_\Omega\rho(x)K_\sigma(x,y)u_K(A_f(x))\dd x, \nonumber\\
 \psi_f^{\rm PR}(y)
 &=\frac{1}{Q_f}\int_\Omega\rho(x)K_\sigma(x,y)
 \omega_K(A_f(x))\dd x, \nonumber\\
 \psi_{\theta,f}(y)
 &=(1-\theta)\phi_f(y)+\theta\psi_f^{\rm PR}(y) \nonumber\\
 &=\int_\Omega\rho(x)K_\sigma(x,y)
 \left[(1-\theta)u_K(A_f(x))
 +\frac{\theta}{Q_f}\omega_K(A_f(x))\right]\dd x,
 \label{eq:theta-gibbs-potential}
\end{align}
where
\begin{equation*}
 u_K(A)=\int_0^1(1-zA)^{K-1}\dd z,
 \qquad
 \omega_K(A)=\frac{1-(1-A)^{K-1}}{KA}.
 \label{eq:common-potential-coefficients}
\end{equation*}

and $\phi_f$ is detailed illustrated in Appendix~\ref{app:foundations}, as it denotes the marginal return from unconditional vote support, while $\psi_f^{\rm PR}$ represents the marginal return from PR seat share, both evaluated by taking unilateral derivatives before imposing the diagonal condition. The spatial gradient difference $\psi_{\theta,f}(y)-\psi_{\theta,f}(y')$ thus measures the net strategic incentive to shift party density from $y'$ to $y$. At any strictly positive stationary profile, the vanishing-flux condition requires $\varepsilon\,\partial_y\log f_\theta^*=\partial_y\psi_{\theta,f_\theta^*}$, which upon spatial integration and unit-mass normalisation yields the Gibbs self-consistency equation:

\begin{equation}
 f^*_\theta(y)=\frac{\exp[\psi_{\theta,f^*_\theta}(y)/\eps]}
 {\int_\Omega\exp[\psi_{\theta,f^*_\theta}(z)/\eps]\dd z}.
 \label{eq:common-gibbs}
\end{equation}

Under the regularity conditions in Appendix~\ref{app:distributional}, this Gibbs mapping admits an even, strictly positive fixed point whose equilibrium width is governed jointly by strategic potential curvature and internal diffusion. We assess the stability of this centrist configuration along a selected stationary branch, recomputing the profile as $K$, $r$, $\eps$, or $\theta$ varies.

To determine whether small inter-party differences grow into sustained separation or decay back to the centre, we perturb this symmetric state along relative-party directions:

\begin{equation}
 f_i=f^*_\theta+\delta z_i g+o(\delta),\qquad
 \sum_i z_i=0,\qquad g(-y)=-g(y).
 \label{eq:standard-perturbation}
\end{equation}

Here, the coefficient vector $z=(z_1,\dots,z_K)$ specifies the relative movement among parties, while the odd deformation $g$ shifts mass across the political centre while strictly preserving each party's unit mass ($\int_\Omega g\dd y = 0$). Owing to party anonymity, the zero-sum constraint $\sum_i z_i=0$ defines a $(K-1)$-dimensional subspace of equivalent divergence directions. Crucially, because the reference profile is even, this deformation perturbs party means and profile skewness while having zero first-order effect on internal party variance.

\subsection{Party number and institutional-payoff weight}

Linearising the marginal strategic potential with respect to these odd deformations reveals that the first-order electoral feedback is governed by the effective weight:

\begin{equation}
 c_{K,\theta}(x)=\frac{\theta}{Q_*}\eta_K(A_*(x))
 +(1-\theta)h_K(A_*(x)),
 \label{eq:theta-feedback-weight}
\end{equation}

where $A_*=A_{f^*_\theta}$, $Q_*=Q_{f^*_\theta}$, and the constituent response kernels are

\begin{align*}
 h_K(A)
 &=\int_0^1z(1-zA)^{K-2}\dd z,
 &h_K(0)&=\frac12, \nonumber\\
 \eta_K(A)
 &=h_K(A)-\frac{(1-A)^{K-2}}{K}
 =\frac{1-(1-A)^{K-2}[1+(K-2)A]}
 {K(K-1)A^2},
 &\eta_K(0)&=\frac{K-2}{2K}.
 \label{eq:feedback-functions}
\end{align*}

with values at $A=0$ defined by continuous extension. Here, $h_K(A)$ captures the direct differentiation response of raw voter mobilisation, whereas $\eta_K(A)$ isolates the residual response retained by seat shares after subtracting the valid-vote normalisation. As derived in Appendix~\ref{app:distributional}, both contributions are strictly non-negative.

In the two-party benchmark ($K=2$), these response functions simplify dramatically:

\begin{equation}
 \eta_2\equiv0,\qquad h_2=\frac12,\qquad
 c_{2,\theta}=\frac{1-\theta}{2}.
 \label{eq:binary-feedback}
\end{equation}

Consequently, under pure seat-share maximisation ($\theta=1$), the reflection-odd feedback vanishes identically in two-party competition ($\eta_2 \equiv 0$), algebraically locking both competitors into the centre. An active feedback channel capable of competing with diffusion emerges only when parties place positive weight on raw vote acquisition ($\theta<1$). For multiparty systems ($K\geq3$), by contrast, $\eta_K(A)>0$ holds on $0<A\leq1$, allowing the institutional seat component itself to fuel divergence. This represents a qualitative shift in the reflection-odd operator: competitor cardinality determines which responses survive valid-vote normalisation, while the payoff weight $\theta$ dictates their relative strength.

\begin{figure}[t]
\centering
\includegraphics[width=\textwidth]{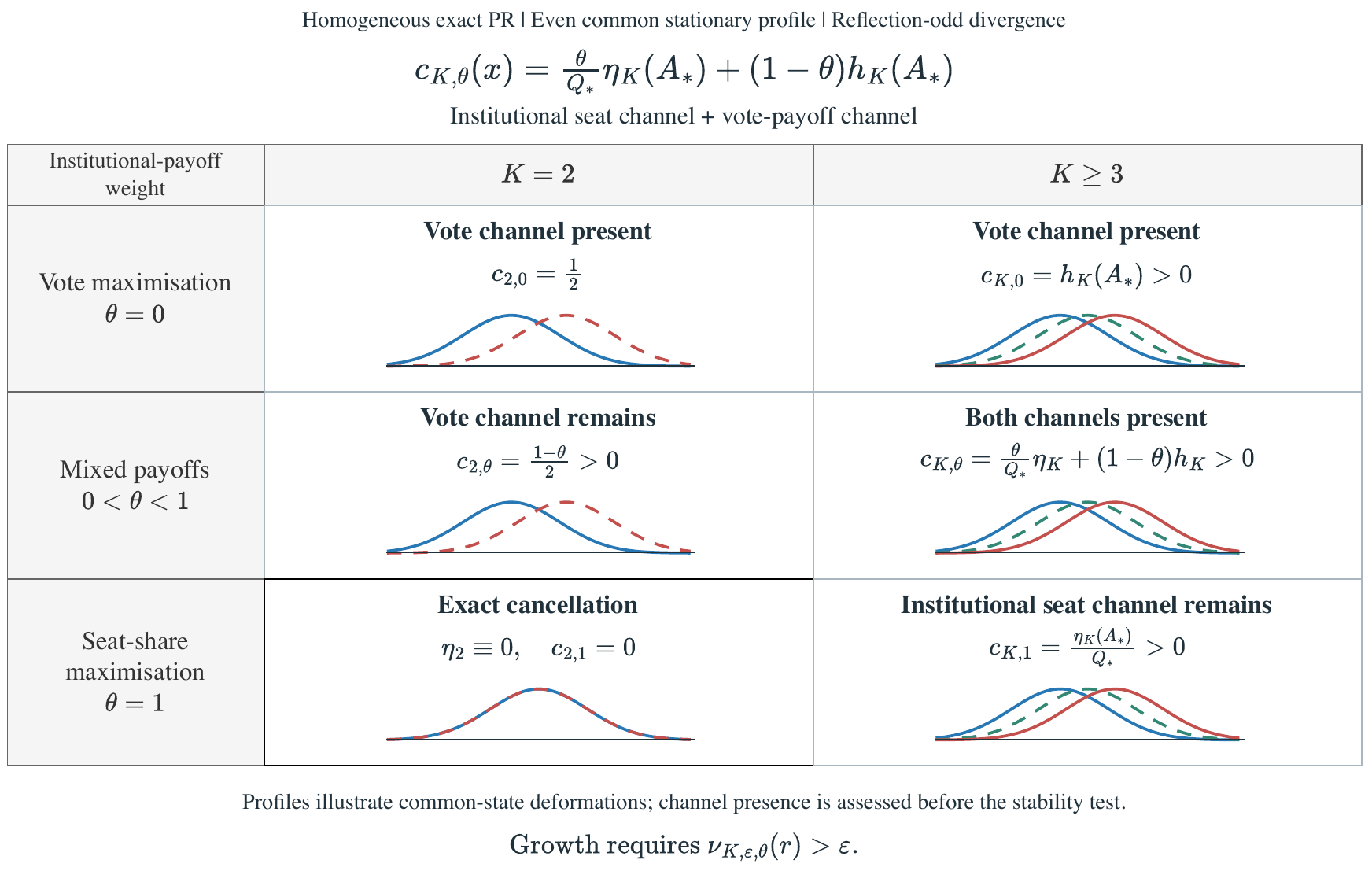}
\caption{\textbf{Party number $\times$ institutional-payoff weight.} Available reflection-odd feedback channels under homogeneous exact PR at an even common stationary profile. Here $A_*=A_*(x)$, with $0<A_*\leq1$, and $Q_*>0$. The binary seat channel cancels exactly, while its vote channel persists for $\theta<1$. For $K\geq3$, the institutional channel survives at $\theta=1$. Density sketches illustrate deformations and the common finite-width state; stability is decided by the gain--diffusion comparison in Figure~\ref{fig:instability-outcome}.}
\label{fig:party-number-payoff}
\end{figure}

Figure~\ref{fig:party-number-payoff} maps these feedback channels across party counts and strategic objectives, establishing the source of differentiation incentives whose ability to destabilise the centrist profile is evaluated below.

\subsection{Feedback against diffusion}

Let $b_g(x)=\int_\Omega K_\sigma(x,y)g(y)\dd y$ denote the first-order acceptance response induced by the ideological deformation $g$. The resulting first-order variation in the strategic potential is generated by the integral operator:

\begin{equation}
 (\mathcal B_{K,\theta}g)(y)
 =\int_\Omega\rho(x)K_\sigma(x,y)c_{K,\theta}(x)b_g(x)\dd x.
 \label{eq:theta-operator}
\end{equation}

The operator $\mathcal B_{K,\theta}$ maps an ideological perturbation directly into its resulting electoral feedback. When restricted to the odd subspace, $\mathcal B_{K,\theta}$ is compact, self-adjoint, and positive semidefinite, satisfying

\begin{equation}
 \langle g,\mathcal B_{K,\theta}g\rangle
 =\int_\Omega\rho(x)c_{K,\theta}(x)b_g(x)^2\dd x.
 \label{eq:theta-operator-quadratic}
\end{equation}

To determine whether this directional feedback overcomes diffusive dispersion, we define the principal strategic gain via the generalised Rayleigh quotient:

\begin{align}
 \nu_{K,\eps,\theta}(r)
 &=\sup_{g\ne0,\ g\ \mathrm{odd}}
 \frac{\langle g,\mathcal B_{K,\theta}g\rangle}
 {\int_\Omega g(y)^2/f^*_\theta(y)\dd y} \nonumber\\
 &=\sup_{g\ne0,\ g\ \mathrm{odd}}
 \frac{\int_\Omega\rho(x)c_{K,\theta}(x)b_g(x)^2\dd x}
 {\int_\Omega g(y)^2/f^*_\theta(y)\dd y}.
 \label{eq:theta-nu}
\end{align}

In this formulation, the numerator measures the total electoral feedback generated by deformation $g$, while the denominator quantifies its amplitude in the weighted $L^2(1/f^*_\theta)$ metric natural to the Fokker--Planck linearisation. Maximising this ratio identifies the optimal mode of differentiation per unit of ideological displacement.

\begin{theorem}[Local differentiation criterion]
\label{thm:distributional-spectral}
At an even stationary profile under homogeneous exact PR, with positive diffusion and the smoothness conditions of Appendix~\ref{app:wellposed}, the reflection-odd party-divergence subspace is linearly stable if $\nu_{K,\eps,\theta}<\eps$ and contains a growing eigenmode if $\nu_{K,\eps,\theta}>\eps$.
\end{theorem}

Theorem~\ref{thm:distributional-spectral} formalises the stability of the political centre as an exact contest between strategic amplification and entropic dispersion. As shown in Appendix~\ref{app:distributional}, the linearised dynamics factor into a positive mobility operator acting on the difference between diffusive and strategic curvatures. At $K=2$ and $\theta=1$, the strategic gain vanishes identically ($\nu=0$), causing odd perturbations to decay exponentially. Once parties value raw mobilisation ($\theta<1$) or compete in a multiparty setting ($K\geq3$), the presence of positive feedback ($\nu>0$) establishes a viable instability channel, triggering divergence whenever this gain exceeds the internal diffusion threshold $\eps$.

Along a selected stationary branch, the balance $\nu=\eps$ identifies the critical voter tolerance $r_c^W(K,\eps,\theta)$ or, at fixed tolerance, the critical institutional weight $\theta_c(r,\eps)$. While the sign of the principal odd eigenvalue $\lambda_{\rm odd}$ strictly tracks $\nu-\eps$, its magnitude depends further on transport mobility and the shape of the reference profile. Consequently, calculating the onset of differentiation and quantifying its temporal growth rate require complementary analyses.

\begin{figure}[t]
\centering
\includegraphics[width=\textwidth]{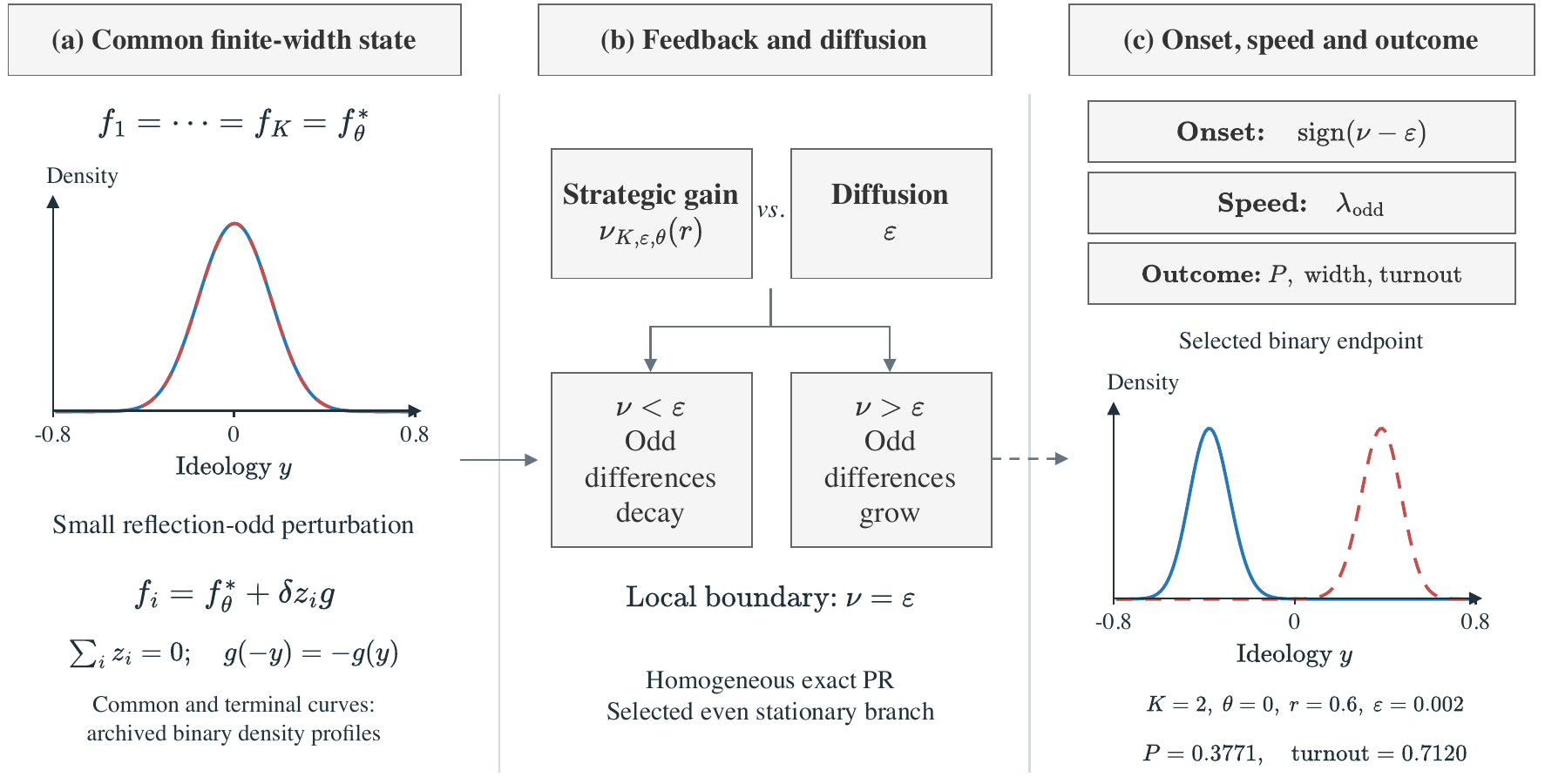}
\caption{\textbf{From local instability to finite-width differentiation.} (a) A common even profile and its reflection-odd party-divergence perturbation. (b) Strategic gain relative to diffusion determines local growth or decay under homogeneous exact PR; equality marks the spectral boundary. (c) Onset, temporal rate and nonlinear outcome are distinct observables. The dashed arrow denotes evaluation by full dynamics. Common and terminal curves use the archived $K=2$, $\theta=0$, $r=0.6$, $\eps=0.002$ run: the terminal profiles have persistent internal width, $P=0.3771$, and turnout rises from $0.6438$ to $0.7120$. Section~\ref{sec:theta-results} reports the endpoint checks.}
\label{fig:instability-outcome}
\end{figure}

Figure~\ref{fig:instability-outcome} bridges this linear criterion with its nonlinear endpoint: the Rayleigh quotient identifies the initial threshold of amplification, the generator eigenvalue governs its growth rate, and the fully coupled transport--diffusion equations determine the terminal separation, internal profile width, and aggregate turnout.

Finally, this finite-width theory smoothly recovers point-platform competition in the vanishing-diffusion limit. Under the regular-branch, centered-well, and transverse-crossing conditions in Appendix~\ref{app:smallnoise}, as $\eps \downarrow 0$ at $\theta=1$, the finite-width critical boundary converges to the classical point-platform threshold, while the leading odd mode approaches an infinitesimal spatial translation of the common profile (see Appendix~\ref{app:pointanalysis} for the Gaussian point benchmark and its large-$K$ asymptotics). This asymptotic connection establishes internal party width as an inertial stabiliser that fundamentally broadens the centrist equilibrium regime.

\FloatBarrier
\section{Phase boundaries and nonlinear differentiation}
\label{sec:theta-results}

To evaluate the theoretical differentiation criterion and trace its nonlinear consequences, our numerical experiments address three sequential questions: 1) where the common centrist profile loses local stability, 2) how rapidly initial perturbations grow, and 3) where the fully coupled transport--diffusion dynamics ultimately settle. We benchmark the model using standard Gaussian voters under homogeneous exact PR, following a selected branch of even Gibbs stationary states as the parameters vary. The baseline analysis fixes internal diffusion at $\eps=0.002$, varies party number across $K\in\{2,3,4,5\}$, and scans the institutional-payoff weight $\theta\in[0,1]$. Computationally, binary growth-rate maps are evaluated over $r\in[0.2,1.1]$ on a $37\times21$ grid, while multiparty root searches cover $r\in[0.5,1.1]$. Appendix~\ref{app:theta-numerics} details the spatial discretisation, stationary residuals, and spectral checks.

\subsection{Objective-dependent phase boundaries}

\begin{figure}[t]
\centering
\includegraphics[width=\textwidth]{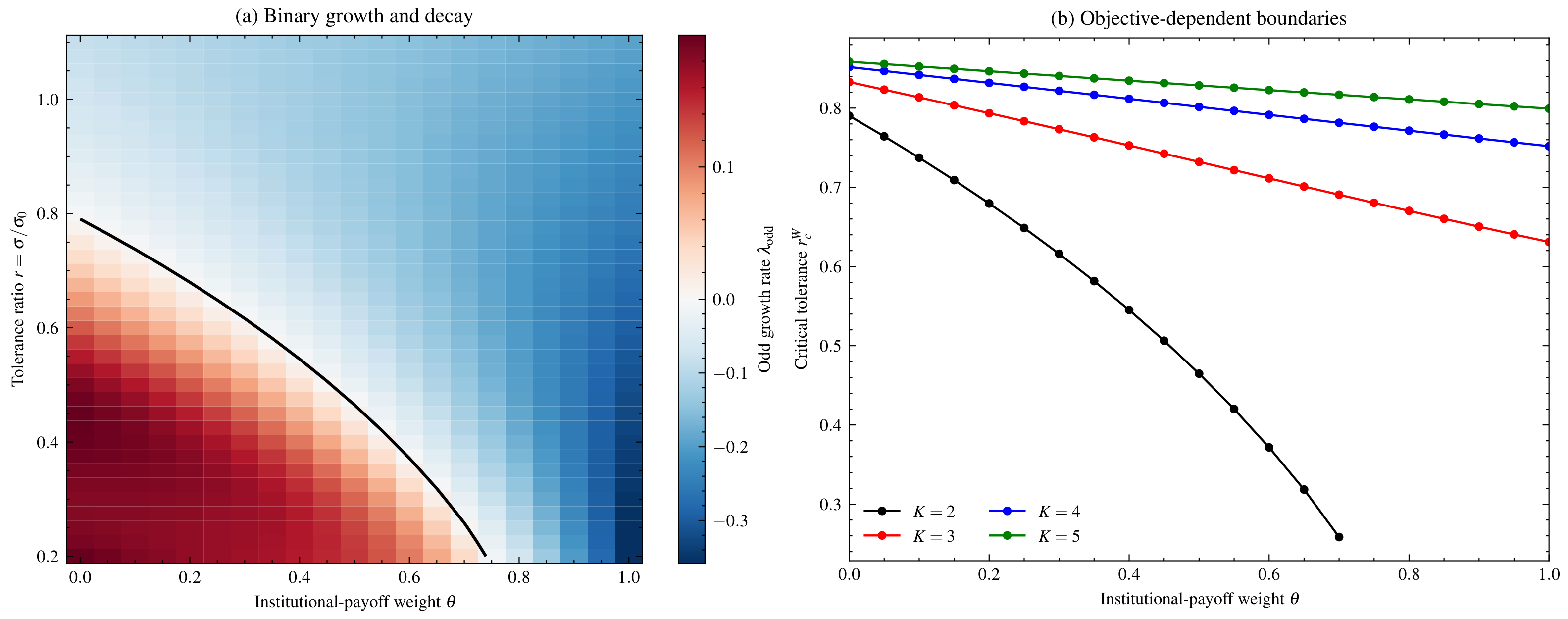}
\caption{Institutional-payoff weight and local differentiation under homogeneous PR, $\eps=0.002$. (a) Binary principal odd growth rate; the black curve marks the zero crossing. (b) Critical tolerance for $K=2,3,4,5$. Points lie on the selected common branch. Binary gaps indicate no detected crossing within $r\in[0.2,1.1]$; multiparty searches use $[0.5,1.1]$. The colours in (a) classify initial odd growth and decay.}
\label{fig:theta-phase}
\end{figure}

Figure~\ref{fig:theta-phase} illustrates how the institutional-payoff weight $\theta$ reshapes the domain of local differentiation by shifting the balance between electoral amplification and diffusive dispersion. In the two-party benchmark ($K=2$), pure vote maximisation ($\theta=0$) yields a critical tolerance of $r_c^W=0.7903$, which contracts to $0.4648$ at an intermediate weight of $\theta=0.5$. Under pure seat-share maximisation ($\theta=1$), the odd strategic gain vanishes entirely, eliminating the instability window. At an intermediate tolerance of $r=0.6$, resolving the neutral boundary yields a critical weight of $\theta_c=0.32364$, such that odd perturbations amplify for $\theta<\theta_c$ and decay for $\theta>\theta_c$ along the selected branch. The phase boundary thus precisely captures the objective weight where directional feedback balances internal diffusion.

Across all four party counts, the critical tolerance curves decrease monotonically with the institutional weight $\theta$ at the sampled values. For multiparty systems ($K=3,4,5$), the instability thresholds under vote maximisation are $0.8329$, $0.8518$, and $0.8585$, whereas under seat-share maximisation they contract to $0.6310$, $0.7519$, and $0.7993$, respectively. Increasing the weight on legislative representation thus systematically narrows the zone of divergence and accentuates the gap between party-number thresholds. This behavior reflects the underlying structure of the feedback operator: while vote mobilisation ($\theta<1$) supplies differentiation incentives across any party count, a stronger emphasis on seat share ($\theta\to 1$) exposes the algebraic cancellation specific to two-party competition.

Importantly, the temporal growth rate responds through a distinct combination of density and mobility effects. At $K=5$ and $r=0.6$, for instance, the principal rate increases from approximately $0.0290$ at $\theta=0$ to $0.0316$ at $\theta=1$, even though the critical tolerance falls and narrows the overall instability window. This occurs because variations in the common density profile and transport mobility can amplify the local drift field at a fixed parameter point. Consequently, the location of the phase boundary cannot be read as a direct proxy for either the speed of initial divergence or the eventual magnitude of party separation.

\subsection{From local growth to separated profiles}

\begin{figure}[t]
\centering
\includegraphics[width=0.96\textwidth]{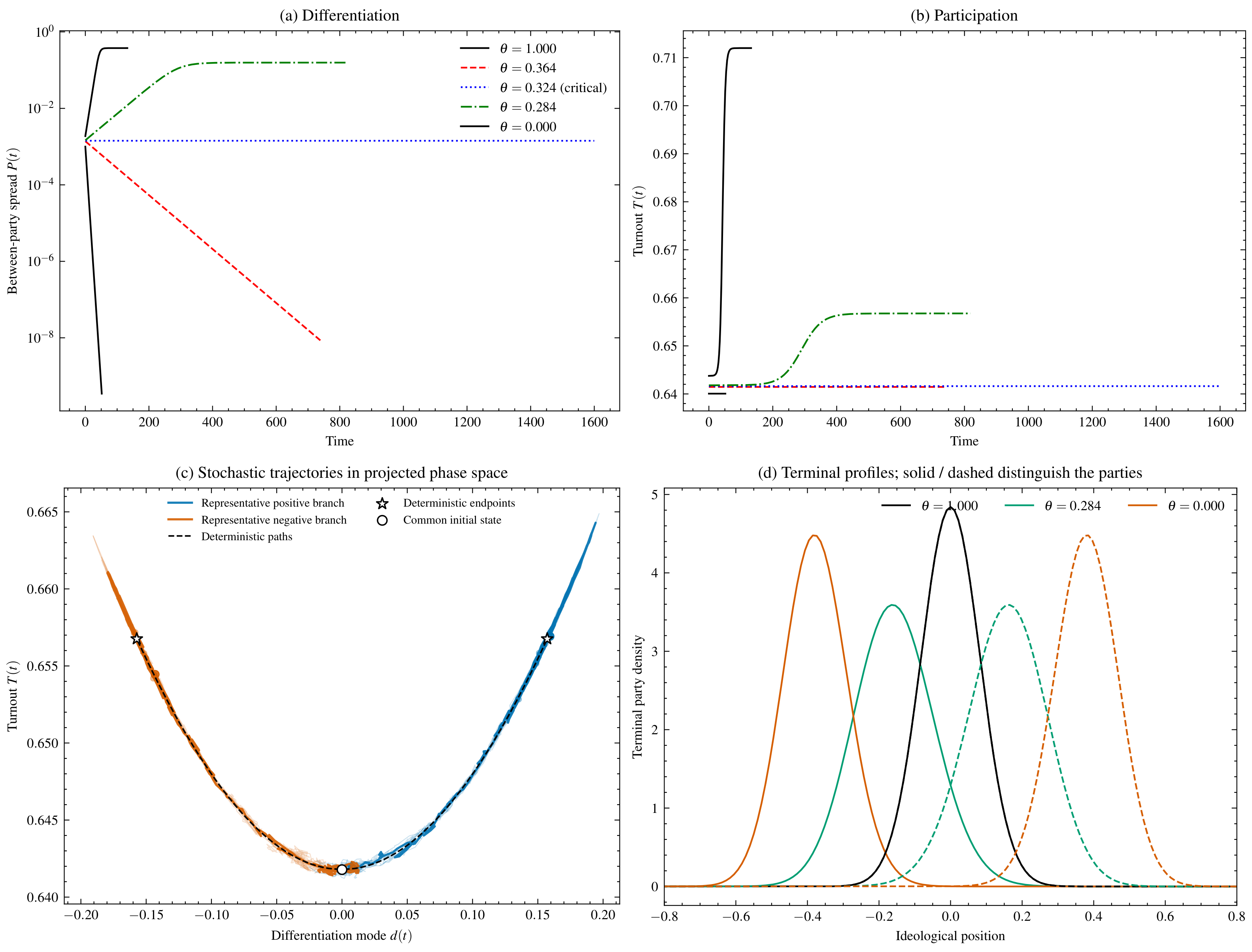}
\caption{Binary dynamics and stochastic phase space at $r=0.6$, $\eps=0.002$. (a)~Between-party spread $P(t)$ and (b)~aggregate turnout $T(t)$ for five values of $\theta$. (c)~Finite-particle trajectories at $\theta\simeq0.284$, projected onto signed differentiation $d=(m_2-m_1)/2$ and turnout. Blue and orange indicate positive and negative terminal branches; seeds 0 and 1 are highlighted among 12 runs with 200 particles per party, each ending at $t=800$. Black dashed curves show the deterministic trajectory and its party-exchanged counterpart; stars mark their endpoints, and the circle marks the common particle initial state. (d)~Terminal profiles for $\theta\in\{1.000,0.284,0.000\}$; solid and dashed lines distinguish the parties. Deterministic runs stop at stationarity tolerance or $t=1600$; critical convergence remains unresolved. Unstable deterministic endpoints pass residual, time-step and asymmetric-restart checks. Numerical methods are given in Appendix~\ref{app:theta-numerics}.}
\label{fig:theta-dynamics}
\end{figure}

To trace the system beyond infinitesimal perturbations, we examine the full nonlinear dynamics by fixing $K=2$ and $r=0.6$, perturbing the common stationary profile with a small, mass-preserving odd deformation that maintains strictly positive initial densities throughout $\Omega$ (specified by Eq.~\eqref{eq:theta-initial} in Appendix~\ref{app:theta-numerics}). The full coupled equations then evolve both party profiles, including their positions, widths, and shapes. Figure~\ref{fig:theta-dynamics} tracks this evolution across pure seat-share maximisation ($\theta=1$), two regimes flanking the critical threshold ($\theta=\theta_c\pm0.04$), the critical state itself, and pure vote maximisation ($\theta=0$).

In line with the linear stability analysis, trajectories in the subcritical regime ($\theta>\theta_c$) decay monotonically back to the centrist profile. In contrast, for $\theta=\theta_c-0.04\simeq0.284$ and $\theta=0$, the system escapes the centrist state and settles into separated stationary configurations with terminal spreads of $P=0.1572$ and $P=0.3771$, respectively. Rather than collapsing into singular point platforms, these terminal distributions retain substantial internal breadth and mutual overlap. Concurrently, aggregate turnout rises from $0.6418$ to $0.6568$ at $\theta=0.284$, and from $0.6438$ to $0.7120$ at $\theta=0$. Ideological separation and voter participation thus advance together: redistributing party platforms not only sharpens ideological distinctiveness but also expands the segment of the electorate that finds at least one party acceptable.

These dynamic endpoints satisfy stringent numerical consistency standards. Across the unstable runs, final drift--diffusion residuals remain below $10^{-9}$ and Gibbs consistency residuals fall below $7\times10^{-9}$. Halving the time step and restarting from asymmetric density tilts recover identical spread $P$ and turnout values to within $2\times10^{-10}$, while initial growth and decay rates match the odd generator spectrum to within two percent away from criticality. Meanwhile, the critical trajectory ($\theta=\theta_c$) hovers close to its initial displacement across the entire horizon ($t=1600$), reflecting the expected vanishing of the spectral gap at the transition boundary.

\subsection{Internal width and diffusion}

\begin{figure}[t]
\centering
\includegraphics[width=\textwidth]{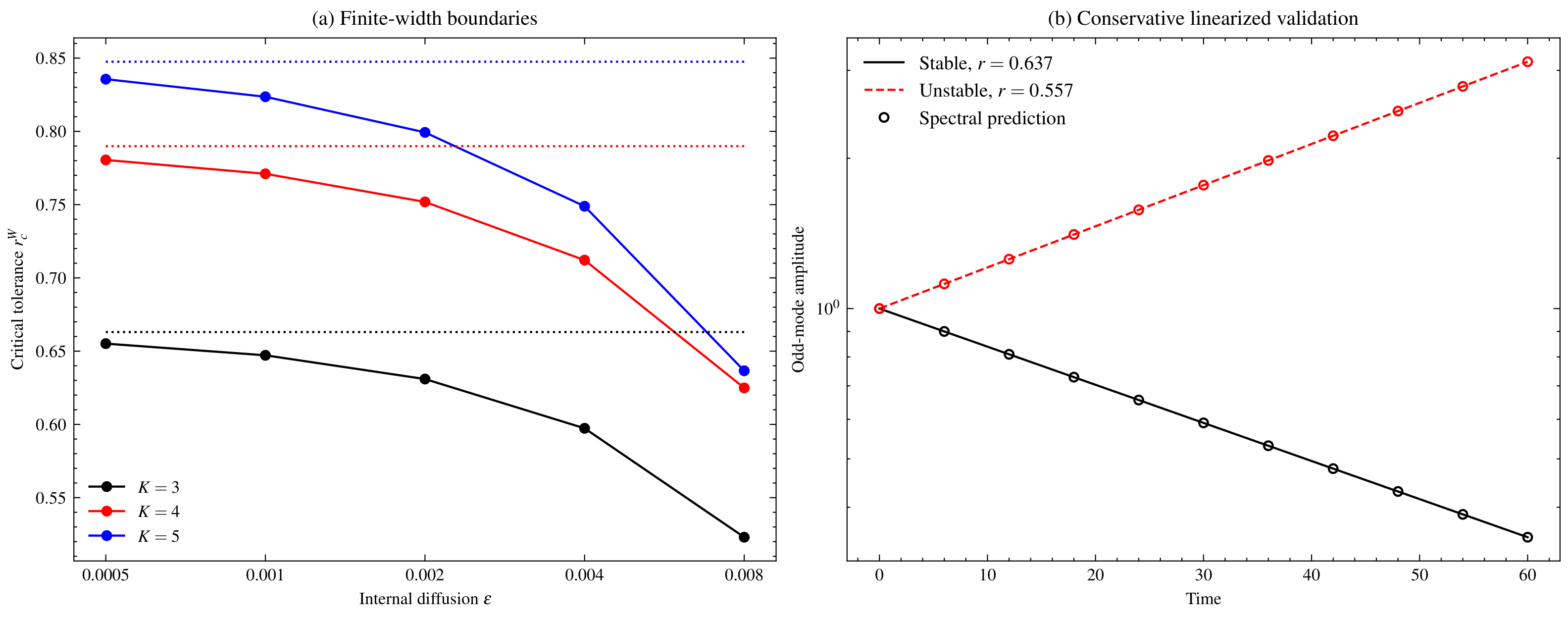}
\caption{Diffusion and local amplification under seat-share maximisation. (a) Critical tolerance versus diffusion for $K=3,4,5$, with point-platform thresholds shown as horizontal references. (b) Conservative linearised dynamics for $K=3$, $\eps=0.004$ on opposite sides of the critical boundary, with theoretical exponential rates. The experiment isolates the growth or decay of a small odd mode. Numerical settings are given in Appendix~\ref{app:numerics}.}
\label{fig:diffusion}
\end{figure}

Internal party dispersion ($\eps$) fundamentally governs both the shape of the baseline profile and the electoral threshold required for differentiation. Figure~\ref{fig:diffusion} displays the finite-width critical boundaries under seat-share maximisation ($\theta=1$) for $K=3,4,5$, alongside a representative local-mode check. At an elevated diffusion of $\eps=0.004$, the critical tolerance ratios are $0.5973$, $0.7121$, and $0.7489$ for $K=3,4,5$; reducing diffusion to $\eps=5\times10^{-4}$ shifts these thresholds upward to $0.6551$, $0.7805$, and $0.8356$. Across all sampled configurations, stronger internal diffusion stabilises the centrist equilibrium over a wider range of voter tolerance, confirming its role as an inertial buffer against centrifugal incentives.

In the small-diffusion regime, the weighted projection of the leading unstable deformation onto the profile gradient $-\partial_yf^*$ exceeds $0.998$ for each party count at the two smallest diffusion levels. This confirms that the critical mode converges to the infinitesimal spatial translation predicted by the small-diffusion asymptotics. Complementary checks under mixed objectives ($\eps\in\{0.001,0.002,0.004\}$) similarly confirm that greater diffusion consistently depresses the critical tolerance threshold. Taken together, these findings demonstrate that internal ideological diversity within a centrist consensus acts as an endogenous stabilizing force, clearly distinguishing intra-party breadth from the inter-party polarisation that materialises once that consensus destabilises.

\section{Electoral institutions and geographic organisation}
\label{sec:institution}

Up to this point, our analysis has examined party adaptation within a single, nationally uniform electorate under proportional representation. In real-world democracies, however, electoral competition is fundamentally governed by distinct institutional allocation rules and the territorial sorting of voters across legislative districts. Spatial theories offer competing expectations on this front, alternatively linking majoritarian or proportional rules to centripetal convergence or centrifugal dispersion~\citep{Cox1990,CalvoHellwig2011}. Empirically, the observed relationship between electoral proportionality and ideological polarisation likewise remains inconclusive and context-dependent~\citep{Dow2011,Ezrow2008}. 

To resolve this ambiguity, the examination of how electoral institutions operate at a distinct stage of the feedback loop is necessary: the conversion of valid votes into parliamentary seats. Analyzing this stage under pure seat-share maximisation ($\theta=1$) allows us to isolate whether cross-system differences in polarisation stem intrinsically from the seat-allocation formula itself, or whether they require geographic heterogeneity across districts to become active. Our model demonstrates that this institutional response cleanly decomposes into two distinct components: a homogeneous responsiveness scale and an institutional curvature activated by territorial vote imbalances.

\subsection{Homogeneous responsiveness}

We first investigate an idealised, geographically homogeneous electorate to evaluate whether electoral rules alone can alter the stability boundary of the centrist equilibrium. For such an electorate, let $R$ denote the seat-allocation map and $v^*=K^{-1}\mathbf{1}$ the symmetric vote vector, where $\mathbf{1}=(1,\ldots,1)^\top$. Under party anonymity, any permutation of party vote shares permutes seat shares identically. Restricted to relative-party divergence directions, the Jacobian at symmetry reduces to a single scalar $g_R$. Assuming positive local responsiveness ($g_R>0$), exact PR yields $g_R=1$, smooth plurality yields $g_R=\beta/K$, and the corresponding value for MMP is derived in Appendix~\ref{app:institutionproof}.

\begin{theorem}[Homogeneous rule equivalence]
\label{thm:effective-temperature}
Under seat-share maximisation, a smooth, party-anonymous rule with $g_R>0$ has reflection-odd divergence operator
\begin{equation}
 \Lop_R^{\mathrm{odd}}(\eps)=g_R\Lop_{\mathrm{PR}}^{\mathrm{odd}}(\eps/g_R)
 \label{eq:effective-temperature}
\end{equation}
at the corresponding homogeneous even stationary state. Matching $\eps/g_R$ therefore gives the same normalised local odd stability boundary.
\end{theorem}

Both operators in Eq.~\eqref{eq:effective-temperature} are evaluated at the identical matched even profile, which is stationary under rule $R$ at diffusion $\varepsilon$ and under PR at an effective diffusion of $\varepsilon/g_R$. Here, the ratio $\eps/g_R$ represents internal ideological dispersion relative to marginal institutional responsiveness. At matched ratios, the centrist state exhibits the identical local stability sign across all anonymous rules, while dynamic growth and decay rates simply scale by $g_R$. In a homogeneous electorate, institutional differences therefore operate purely as a response velocity rather than altering the bifurcation boundary. This equivalence holds because reflection symmetry removes the first-order vote response of an odd deformation, which in turn eliminates its contraction with institutional curvature (see Appendix~\ref{app:institutionproof}). Crucially, this benchmark reveals that electoral institutions cannot natively reshape the stability boundary of a common centre in the absence of territorial variation.

\subsection{Geographic curvature}

Because homogeneous electorates neutralise institutional non-linearities, the capacity of electoral rules to govern party divergence depends centrally on the territorial distribution of political support~\citep{Rodden2010,Linzer2012,CalvoRodden2015}. To analyze this channel, consider equal-weight reflected district pairs, $\rho_{\bar d}(x)=\rho_d(-x)$, treated symmetrically by the electoral rule. When parties undergo a reflection-odd displacement, they induce equal and opposite first-order vote responses in paired districts. Consequently, while their national first-order vote changes cancel out, the quadratic products of these district-level responses preserve their sign.

The composite mapping connecting party platforms to legislative seats is given by $y\to v(y)\to\mathcal S^R(v(y))$. Unilateral platform shifts first perturb district support and turnout, modifying local valid-vote shares; the seat-map gradient then translates these shifts into representation. Differentiating this composite map to second order separates the response into two distinct mechanisms: the curvature of the behavioural vote response weighted by marginal seat gains, and the intrinsic curvature (Hessian) of the seat-allocation rule contracting along district vote responses (see Appendix~\ref{app:institutionproof} for explicit chain rules).

This second term—the geographic institutional channel—survives reflection whenever the seat-map Hessian has a non-zero contraction along district response vectors. In two-party competition ($K=2$), however, party anonymity and seat-share conservation ($\mathcal{S}_1^R + \mathcal{S}_2^R = 1$) impose an additional identity: the seat Hessian vanishes identically on the district vote tangent space at the tie. Reversing binary labels flips all district vote imbalances while mapping each seat share to its complement, forcing the institutional quadratic term at the symmetric state to zero. Stability in the binary setting is therefore determined solely by the restoring behavioural vote curvature. In multiparty systems ($K\geq3$), by contrast, this cancellation breaks down, and geography directly activates the local curvature of the seat map.

\subsection{Comparisons at matched national density}

To establish that cross-rule differences in polarisation arise from district organisation rather than confounding differences in the national electorate, each heterogeneous district configuration must be benchmarked against a homogeneous control sharing the exact same national preference density $\bar{\rho}_g$. Both configurations in each pair use the same electoral rule, parameters, and initial platform positions. We test this across $K=5$ parties, nine equal-weight districts, tolerance $\sigma=0.65$, and five initialisations, integrating point-platform dynamics to $t=42$. The geography parameter $g$ scales district means across standard normal quantiles while fixing within-district standard deviations at $\sqrt{1-g^2}$ on $[-4,4]$ (detailed in Appendix~\ref{app:matched-geography}).

\begin{figure}[t]
\centering
\includegraphics[width=\textwidth]{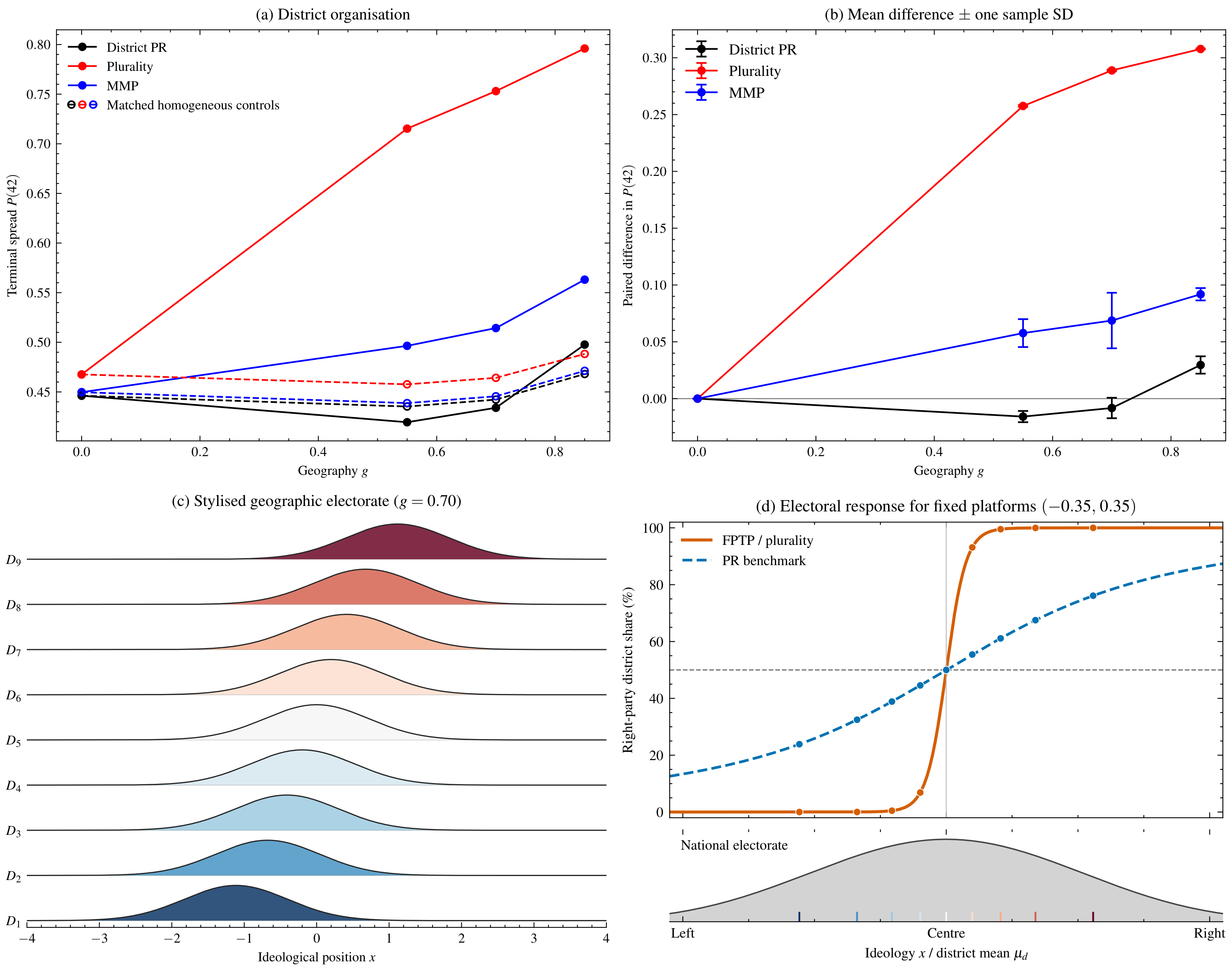}
\caption{\textbf{District organisation and institutional curvature at matched national density ($\theta=1$).} 
(a)~Terminal mean spread $P(42)$ across geography $g$ for five-party systems (solid) and matched homogeneous controls (dashed). (b)~Paired differences ($P_g - P_g^{\mathrm{hom}}$) with one-standard-deviation error bars across five matched seeds, isolating the net effect of district organisation. (c)~Spatial voter distributions across nine districts ($D_1$--$D_9$) at $g = 0.70$. (d)~Electoral responsiveness for fixed platforms at $y = \pm 0.35$, highlighting the steep institutional curvature of plurality compared to the buffered PR benchmark. Bottom marginal depicts the national electorate with colored ticks indicating district means $\mu_d$. Parameters are $K=5$, $D=9$, $\sigma=0.65$, and $t=42$; PR and MMP include the smooth five-percent threshold.}
\label{fig:matched-geography}
\end{figure}

When districts are homogeneous ($g=0$), outcomes between paired configurations coincide identically. As geographic sorting intensifies to $g=0.70$, district organisation markedly amplifies divergence, increasing terminal party spread by $0.289$ under single-member plurality and $0.069$ under MMP, while producing a negligible difference of $-0.008$ under PR. Correspondingly, the plurality--PR spread gap expands to $0.319$ with district sorting, compared to just $0.022$ in the matched homogeneous control. As shown in Figure~\ref{fig:matched-geography}, this demonstrates that the pronounced divergence generated by plurality rule is not an inherent property of majoritarian voting in isolation, but an endogenous consequence of district organisation interacting with the winner-take-all curvature of the seat map.

In sharp contrast, the analogous two-party experiments at $\theta=1$ converge toward coincident centrist platforms across all tested geographic environments. The largest observed terminal spread is merely $P(42)=9.41\times10^{-7}$, which decays below $8.0\times10^{-12}$ when extended to $t=84$. Local divergence rates remain strictly negative across all three rules (Table~\ref{tab:binary-local-rates} in Appendix~\ref{app:binary-geography}). This negative result confirms that the algebraic cancellation of institutional curvature in two-party competition holds regardless of district heterogeneity, insulating two-party systems from the geographic polarisation engine that destabilises multiparty competition.

\section{Discussion and conclusion}
\label{sec:discussion}

The dynamical framework developed in this paper establishes that the stability of a common political centre is fundamentally governed by the balance between directional electoral feedback and internal ideological dispersion. A centrist consensus persists so long as internal party diffusion absorbs small ideological departures; it breaks down when the electoral payoff structure amplifies reflection-odd perturbations into sustained separation. Rather than discarding the logic of centripetal competition, this perspective situates the classic Downsian equilibrium within a broader parameter space, demonstrating that convergence and polarisation represent dual, coexisting regimes of a single underlying competitive process.

The transition between these regimes hinges decisively on the interplay between party cardinality and strategic objectives. Under pure seat-share maximisation ($\theta = 1$), a two-party system enforces strict centripetal convergence: because valid-vote shares are strictly conserved, unilateral ideological shifts yield an exact algebraic cancellation of the odd response operator, locking both competitors into the centre. This centrist lock is broken only when parties value raw voter mobilisation ($\theta < 1$), where the prospect of activating disaffected abstainers on the ideological flanks generates an active differentiation channel. In multiparty settings ($K \ge 3$), however, the strategic landscape shifts fundamentally: the physical crowding of the centre dilutes its marginal electoral returns, enabling parties to secure parliamentary advantage by diverging outward even under strict seat-seeking incentives. Across all configurations, internal party dispersion ($\varepsilon$) serves as an inertial stabiliser, systematically expanding the range of voter tolerance over which the centrist consensus remains robust.

Electoral institutions modulate this feedback by governing how localized voter support translates into parliamentary power. In geographically homogeneous electorates, institutional rules differ solely in their marginal responsiveness, scaling the temporal speed of adaptation without altering the underlying bifurcation boundary. In contrast, geographic variation across districts activates the intrinsic curvature of the seat-allocation rule. Under single-member plurality, the spatial concentration of voter preferences interacts with district thresholds to sharply expand the zone of divergence, turning geographic sorting into a powerful engine of multiparty polarisation. Proportional representation, by mitigating these localized non-linearities, functions as an institutional buffer that dampens centrifugal forces.

By conceptualizing political parties as continuous probability distributions rather than point platforms, the model formally uncouples two dimensions of political competition that are frequently conflated: inter-party polarisation and intra-party heterogeneity. An unstable centrist equilibrium does not imply a collapse into ideological extremism or rigid dogmatism; rather, parties can achieve sustained spatial separation while retaining substantial internal variance and mutual overlap. Furthermore, because participation is endogenous, ideological differentiation reshapes not only the alignment of party platforms but also the overall size and compositional makeup of the active electorate.

These results offer a principled foundation for examining more complex competitive environments. Natural extensions include generalizing the ideological domain to multidimensional policy spaces—where parties may diverge on socio-cultural issues while converging on economic policy—and incorporating dynamic voter feedback, in which evolving party movements actively reshape the underlying distribution of voter preferences over time. Ultimately, our findings demonstrate that party differentiation is neither an inevitable response to an already poarlised electorate nor an institutional accident, but an endogenous consequence of how strategic objectives, competitor number, internal distribution profiles, and vote-to-seat rules structure the returns to political competition.

\section*{Declaration of competing interest}

The authors declare that they have no known competing financial interests or personal relationships that could have appeared to influence the work reported in this paper.

\section*{Declaration of generative AI and AI-assisted technologies in the manuscript preparation process}
During the preparation of this work the authors used ChatGPT for grammar checking and language polishing. After using this tool/service, the authors reviewed and edited the content as needed and take full responsibility for the content of the published article.

\section*{Acknowledgment}
This work is supported by the European Research Council (ERC) under the European Union's Horizon 2020 research and innovation program (grant agreement No. 101002240). We also thank Prof. Thomas Chadefaux from Trinity College Dublin and Prof. Yuhui Deng from Beijing Normal-Hong Kong Baptist University for helpful comments on the manuscript.

\bibliographystyle{unsrt}
\bibliography{references}

\appendix
\section{Choice operator, first variation, and well-posedness}
\label{app:foundations}

The exact choice identity and its first variation connect individual acceptance to the combined strategic potential. Both payoff components generate bounded, measure-Lipschitz kernel potentials. Appendix~\ref{app:distributional} treats arbitrary $\theta$; the small-diffusion and institutional results concern $\theta=1$.

\label{app:choiceproof}
To obtain Eq.~\eqref{eq:choice}, condition on party $i$ being acceptable ($B_i=1$). The voter selects uniformly among party $i$ and the other accepted alternatives, giving the conditional probability
\begin{equation}
 \E\!\left[\frac{1}{1+\sum_{j\neq i}B_j}\right].
\end{equation}
Using the integral identity $(n+1)^{-1}=\int_0^1z^n\dd z$ and exploiting the mutual independence of the acceptance indicators across parties, we obtain
\begin{align}
 p_i
 &=a_i\int_0^1\E\!\left[z^{\sum_{j\neq i}B_j}\right]\dd z\\
 &=a_i\int_0^1\prod_{j\neq i}\bigl[(1-a_j)+za_j\bigr]\dd z,
\end{align}
which gives Eq.~\eqref{eq:choice}. Abstention occurs precisely when every $B_j=0$, giving $p_0=\prod_j(1-a_j)$. Selection and abstention exhaust the outcomes, so $\sum_i p_i=1-p_0$. The integrand has degree at most $K-1$; an $n$-node Gauss--Legendre rule is exact for $2n-1\geq K-1$, giving the stated $\lceil K/2\rceil$ nodes.

\label{app:firstvariationproof}
For the marginal seat potential, define $M_{\ell i}=\partial p_\ell/\partial a_i$ and $R_{\ell d}^{(i)}=\partial\mathcal S_i^R/\partial v_{\ell d}$. Here $i$ identifies the adapting party, $\ell$ the party receiving votes and $d$ the district. The coefficient $M$ converts acceptance into voter choice; $R$ converts district valid votes into seats, including the rule's weights. The own-party response $M_{ii}$ is non-negative and competitor responses $M_{\ell i}$ are non-positive. Perturb party $i$ by a zero-mass signed measure $\eta_i$ and define
\begin{equation}
 \delta a_i(x)=\int_\Omega K_\sigma(x,y)\dd\eta_i(y).
\end{equation}
Holding all competing party profiles fixed, the perturbation alters only $a_i$, producing the choice response $\delta p_\ell=M_{\ell i}\delta a_i$. Differentiating Eq.~\eqref{eq:choice} under the integral sign gives
\begin{equation}
 M_{ii}(a)=\int_0^1\prod_{j\neq i}[1-(1-z)a_j]\dd z,
 \qquad
 M_{\ell i}(a)=a_\ell\int_0^1\partial_{a_i}
 \!\left(\prod_{j\neq\ell}[1-(1-z)a_j]\right)\dd z
 \quad(\ell\neq i).
 \label{eq:Mentries}
\end{equation}
Integrating these individual choice responses over the district voter distribution yields the variation in raw voter support:
\begin{equation}
 \delta q_{\ell d}=\int\rho_d(x)M_{\ell i}(a(x))\delta a_i(x)\dd x.
\end{equation}
Differentiating the quotient $v_{\ell d}=q_{\ell d}/T_d$ with respect to party profiles
\begin{equation}
 \delta v_{\ell d}=\frac{1}{T_d}
 \left(\delta q_{\ell d}-v_{\ell d}\sum_m\delta q_{md}\right).
\end{equation}
Substitute into $\delta\mathcal S_i^R=\sum_{d,\ell}R_{\ell d}^{(i)}\delta v_{\ell d}$ and apply Fubini's theorem. The coefficient of $\dd\eta_i(y)$ is
\begin{equation}
 \psi_i^R(y)=\sum_{d,\ell}\frac{R_{\ell d}^{(i)}}{T_d}
 \int\rho_d(x)\left[M_{\ell i}(a(x))-v_{\ell d}\sum_mM_{mi}(a(x))\right]
 K_\sigma(x,y)\dd x.
 \label{eq:firstvariation}
\end{equation}
The kernel converts redistribution at $y$ into acceptance at $x$. Within the brackets, the direct support response is corrected for the change in total turnout. Division by $T_d$ expresses this in valid-vote units, and $R$ converts it into representation. All coefficients are evaluated at the current party state. The support potential is
\begin{equation}
 \phi_i(y)=\sum_dw_d\int\rho_d(x)M_{ii}(a(x))K_\sigma(x,y)\dd x.
 \label{eq:support-potential}
\end{equation}
Their objective-weighted combination gives Eq.~\eqref{eq:mixed-potential}.

\label{app:wellposed}
Existence, uniqueness and parabolic regularity for Eq.~\eqref{eq:pde} follow under the following conditions. Assume that $\Omega$ is compact and convex with a $C^2$ boundary. Let each voter density satisfy $\rho_d\in L^\infty$. Assume that $K_\sigma$ is strictly positive and $C^3$. Let each seat map be $C^2$ on the relevant compact vote domain, and let $\eps_i\geq\underline\eps>0$. Strict kernel positivity gives a uniform lower bound on turnout. If $K_\sigma\geq c>0$, then $a_i(x)\geq c$ and
\begin{equation}
 T_d\geq1-(1-c)^K>0.
\end{equation}
Thus $T_d$ is uniformly bounded away from zero. The choice map and its first two derivatives are bounded on $[0,1]^K$, ensuring that valid-vote normalisation introduces no algebraic singularities on the state space. Under these uniform bounds and Eqs.~\eqref{eq:firstvariation}--\eqref{eq:support-potential}, the drift field $b_i=\nabla\psi_i^{R,\theta}$ is uniformly bounded and satisfies a global Lipschitz continuity condition in $y$ and in the party measure profile under the Wasserstein-1 metric $W_1$:
\begin{equation}
 \lVert b_i(\cdot;\bm\mu)-b_i(\cdot;\bm\nu)\rVert_\infty
 \leq C\sum_jW_1(\mu_j,\nu_j).
 \label{eq:drift-lipschitz}
\end{equation}
Fix a continuous candidate path $\bm m(t)$. Let $\Phi(\bm m)$ denote the law of the reflected diffusion
\begin{equation}
 \dd X_i=b_i(X_i;\bm m(t))\dd t+\sqrt{2\eps_i}\dd B_i-n(X_i)\dd L_i.
\end{equation}
The convex Skorokhod problem with Lipschitz drift has a unique strong solution~\citep{LionsSznitman1984}. Coupling the trajectories for two candidate paths $\bm m$ and $\bm n$ driven by the same Brownian motion, the non-expansiveness of normal boundary reflection combined with Gronwall's inequality implies
\begin{equation}
 \sup_{t\leq\tau}\sum_iW_1(\Phi_i(\bm m)(t), \Phi_i(\bm n)(t))
 \leq C\tau e^{C\tau}
 \sup_{t\leq\tau}\sum_iW_1(m_i(t), n_i(t)).
\end{equation}
For sufficiently small $\tau$, the map $\Phi$ is a strict contraction. Repeating the argument on consecutive intervals gives a unique solution on every finite horizon for initial probability measures. The reflected diffusion preserves probability mass and non-negativity. For initial densities $f_i(0)\in L^2(\Omega)$, standard uniformly parabolic estimates give
\begin{equation}
 f_i\in L^2(0,T;H^1(\Omega)),
 \qquad
 \partial_tf_i\in L^2(0,T;H^{-1}(\Omega)).
\end{equation}
For general initial probability measures, these estimates hold on $[\tau,T]$ for every $0<\tau<T$.

\section{Symmetric equilibrium and the exact divergence spectrum}
\label{app:distributional}

Permutation symmetry reduces the divergence spectrum to an operator on ideological profiles. For a common density $f$, define
\begin{equation}
 A_f(x)=\int K_\sigma(x,y)f(y)\dd y,\qquad
 Q_f=\int\rho(x)[1-(1-A_f(x))^K]\dd x.
 \label{eq:AfQf}
\end{equation}
The coefficient functions needed on the diagonal are
\begin{align}
 \omega_K(A)&=\frac{1-(1-A)^{K-1}}{KA},& \omega_K(0)&=\frac{K-1}{K},
 \label{eq:omega}\\
 u_K(A)&=\int_0^1(1-zA)^{K-1}\dd z,&
 h_K(A)&=\int_0^1z(1-zA)^{K-2}\dd z,
 \label{eq:theta-raw-coefficients}\\
 \eta_K(A)&=\frac{1-(1-A)^{K-2}[1+(K-2)A]}{K(K-1)A^2},&
 \eta_K(0)&=\frac{K-2}{2K}.
 \label{eq:chi-eta}
\end{align}
They satisfy $h_K(A)=\eta_K(A)+(1-A)^{K-2}/K$. At diagonal attraction $A$, each party receives $[1-(1-A)^K]/K$. Differentiating choice and valid-vote normalisation gives the seat-share potential
\begin{equation}
 \psi_f(y)=\frac1{Q_f}\int\rho(x)\omega_K(A_f(x))K_\sigma(x,y)\dd x,
 \label{eq:diagonal-potential}
\end{equation}
up to an additive constant. The raw-support derivative contributes $u_K$, so the general potential is
\begin{equation}
 \psi_{\theta,f}(y)=\int\rho(x)K_\sigma(x,y)
 \left[\frac{\theta}{Q_f}\omega_K(A_f(x))+(1-\theta)u_K(A_f(x))\right]\dd x.
 \label{eq:theta-gibbs-potential-app}
\end{equation}
The Gibbs map is
\begin{equation}
 \mathcal G_{\eps,\theta}(f)=\frac{e^{\psi_{\theta,f}/\eps}}
 {\int e^{\psi_{\theta,f}(z)/\eps}\dd z}.
 \label{eq:gibbs-map}
\end{equation}
For $\theta=1$ we write $\mathcal G_\eps$ and $f^*_{K,r,\eps}$.

Let $\mathcal P_{\mathrm{even}}(\Omega)$ be the weakly compact convex set of even probability measures. The map in Eq.~\eqref{eq:gibbs-map} sends this set into strictly positive, even, continuous densities. Kernel continuity makes $A_f$, $Q_f$ and $\psi_{\theta,f}$ continuous under weak convergence. Strict kernel positivity keeps $Q_f$ uniformly away from zero. The map is continuous with relatively compact image in $C(\Omega)$, so Schauder's theorem gives an even fixed point. It has positive internal variance. The argument permits multiple fixed points and is independent of convergence of any numerical iteration.

Define $\Tspace=\{z\in\R^K:\1^\top z=0\}$, which has dimension $K-1$. Linearise along $f_i=f^*+\delta z_i g+o(\delta)$ with $z\in\Tspace$, and write $b=b_g$. The first-order variation of total turnout vanishes at the diagonal because $\sum_jz_j=0$. For the seat component, Eq.~\eqref{eq:firstvariation} and differentiation of the choice rule give
\begin{equation}
 \delta q_i=z_i\ell[g],
 \qquad
 \ell[g]=\int\rho(x)\chi_K(A_*(x))b(x)\dd x,
 \label{eq:ell}
\end{equation}
where
\begin{equation}
 \chi_K(A)
 =\int_0^1(1-uA)^{K-1}\dd u
 +A\int_0^1u(1-uA)^{K-2}\dd u
 =\frac{1-(1-A)^{K-1}}{(K-1)A}.
\end{equation}
Since turnout variation vanishes in this sector, $\delta S_i=z_i\ell[g]/Q_*$. The remaining coefficient in the marginal seat potential is
\begin{align}
 &\int_0^1u(1-uA)^{K-2}\dd u-\frac1K(1-A)^{K-2}\\
 &\hspace{25mm}=\frac{1-(1-A)^{K-2}[1+(K-2)A]}
 {K(K-1)A^2}=\eta_K(A).
\end{align}
Substitution gives
\begin{equation}
 \delta\psi_i=z_i\,\mathcal E_{K,f^*}[g],
 \label{eq:standard-linearisation}
\end{equation}
with
\begin{align}
 \mathcal E_{K,f^*}[g](y)
 =&\frac1{Q_*}\int\rho(x)K_\sigma(x,y)
 \eta_K(A_*(x))b(x)\dd x
 \nonumber\\
 &-\frac{\ell[g]}{Q_*^2}
 \int\rho(x)(1-A_*(x))^{K-1}K_\sigma(x,y)\dd x.
 \label{eq:Eoperator}
\end{align}
Thus the divergence representation appears with multiplicity $K-1$. Every party-label direction carries the same operator on ideological deformations. Taylor expansion gives the continuous values of $\chi_K$ and $\eta_K$ at $A=0$. For $K=2$, the numerator of $\eta_K$ vanishes identically. For $K\geq3$, strict decrease of $m\log(1-A)+\log(1+mA)$ implies $\eta_K(A)>0$ on $0<A\leq1$, where $m=K-2$.

For the raw-support component, $M_{ii}=\int_0^1\prod_{j\ne i}(1-ua_j)\dd u$. Differentiating along $\delta a_j=z_jb$ gives $z_i h_K(A_*)b$, using $\sum_{j\ne i}z_j=-z_i$. The two responses combine with the weights in Eq.~\eqref{eq:payoff}.

Suppose the reference profile, electorate and kernel are reflection symmetric and $g$ is odd. Then $b_g$ is odd and $\chi_K(A_*)$ is even, so $\ell[g]=0$. At the corresponding mixed-objective profile, the resulting operator is
\begin{equation}
 (\mathcal B_{K,\theta}g)(y)=\int\rho(x)K_\sigma(x,y)c_{K,\theta}(x)b_g(x)\dd x.
 \label{eq:theta-operator-app}
\end{equation}
For seat-share maximisation this is
\begin{equation}
 (\Bop_{K,f^*}g)(y)=Q_*^{-1}\int\rho(x)K_\sigma(x,y)\eta_K(A_*(x))b_g(x)\dd x.
 \label{eq:Boperator}
\end{equation}
The kernel is symmetric, continuous and positive semidefinite, since
\begin{equation}
 \langle g,\mathcal B_{K,\theta}g\rangle=\int\rho(x)c_{K,\theta}(x)b_g(x)^2\dd x\geq0.
 \label{eq:theta-quadratic}
\end{equation}
Thus the operator is compact and self-adjoint. Linearising Eq.~\eqref{eq:pde} at the zero-current state and using
\begin{equation}
 \partial_y\psi^*=\eps\,\partial_y\log f^*
\end{equation}
gives
\begin{equation}
 \partial_tg=\eps\partial_y\!\left[f^*\partial_y(g/f^*)\right]
 -\partial_y\!\left[f^*\partial_y(\mathcal B_{K,\theta}g)\right]
 \label{eq:odd-linearized-pde}
\end{equation}
with linearised no-flux boundary conditions. On the zero-mass odd space, define the positive mobility operator
\begin{equation}
 \mathcal M_{f^*}u=-\partial_y(f^*\partial_yu)
\end{equation}
with Neumann flux boundary conditions. Then
\begin{equation}
 \partial_tg=-\mathcal M_{f^*}
 \left(\eps\frac{g}{f^*}-\mathcal B_{K,\theta}g\right).
 \label{eq:MHfactor}
\end{equation}
The bracketed Hessian has quadratic form
\begin{equation}
 \mathcal Q[g]
 =\eps\int\frac{g^2}{f^*}\dd y-\langle g,\mathcal B_{K,\theta}g\rangle.
 \label{eq:Hessian-form}
\end{equation}
The compact generalised eigenproblem
\begin{equation}
 \mathcal B_{K,\theta}g=\nu\frac{g}{f^*}
\end{equation}
has largest eigenvalue $\nu_{K,\eps,\theta}(r)$ from Eq.~\eqref{eq:theta-nu}. If $\nu<\eps$, Eq.~\eqref{eq:Hessian-form} is positive definite and the linearised generator has negative spectrum. If $\nu>\eps$, the Hessian has a negative direction. Conjugation by $\mathcal M_{f^*}^{1/2}$ and the min--max principle give a positive growth eigenvalue, proving Theorem~\ref{thm:distributional-spectral}. At $K=2,\theta=1$, the strategic operator vanishes and a weighted Poincar\'e inequality gives exponential odd decay. For the seat-share objective considered in the limit below, write
\begin{equation}
 \nu_{K,\eps}(r):=\nu_{K,\eps,1}(r)
 =\sup_{g\ne0,\ g\ \mathrm{odd}}\frac{Q_*^{-1}\int\rho\eta_K(A_*)b_g^2\dd x}{\int g^2/f^*\dd y}.
 \label{eq:nu}
\end{equation}

\section{Small-dispersion concentration and recovery of the point-platform boundary}
\label{app:smallnoise}

\begin{theorem}[Conditional small-diffusion limit]
\label{thm:smallnoise}
Fix $K\geq3$ and $\theta=1$. Under Assumptions~\ref{ass:centered-well} and~\ref{ass:regular-branch}, suppose the point threshold $r_c(K)$ is a transverse crossing. For sufficiently small positive $\eps$, the centred branch has odd critical ratios near $r_c(K)$. Any such selection within a sufficiently small fixed neighbourhood satisfies
\begin{equation}
 r_c^W(K,\eps)\to r_c(K),\qquad
 \nu_{K,\eps}/\eps\to C_K/\kappa_K,\qquad
 g_{\rm crit}\sim-\partial_yf^*_{K,r,\eps}.
 \label{eq:smallnoise-results}
\end{equation}
The mode is understood up to normalisation in the concentration scale. Its ideological shape is common to the $K-1$ equivalent party-label directions.
\end{theorem}

The small-dispersion connection follows from concentration of the centred Gibbs profile and the resulting leading term of the odd-sector Rayleigh quotient. For the standard Gaussian benchmark, let the centred point-platform state be $f_0=\delta_0$. Define $s_r(x)=K_r(x,0)$ and
\begin{align}
 Q_0(r)&=\int\rho(x)[1-(1-s_r(x))^K]\dd x,\\
 \psi_0(y;r)&=\frac1{Q_0(r)}\int\rho(x)\omega_K(s_r(x))K_r(x,y)\dd x.
 \label{eq:point-diagonal-potential}
\end{align}
The continuation argument uses the following local conditions.

\begin{assumption}[{Centrist strategic concavity}]
\label{ass:centered-well}
For $r$ in a neighbourhood of $r_c(K)$, $\psi_0(\cdot;r)$ has a unique global maximum at zero and
\begin{equation}
 \kappa_K(r)=-\psi_0''(0;r)>0.
\end{equation}
The kernel is $C^4$ in its second argument. The unique maximum and its exterior gap are uniform over that neighbourhood.
\end{assumption}

\begin{assumption}[Regular centred branch]
\label{ass:regular-branch}
For every sufficiently small $\eps>0$, the centred fixed points considered below form a continuous branch in $r$ on a fixed neighbourhood of $r_c(K)$. At each point of this branch, $I-D\mathcal G_\eps(f^*_{K,r,\eps})$ is invertible on $X_0=\{h\in C_{\mathrm{even}}(\Omega):\int h=0\}$.
\end{assumption}

The invertibility in Assumption~\ref{ass:regular-branch} is a non-degeneracy condition on the Gibbs fixed point. It permits local continuation of the selected stationary profile in $r$ at each fixed positive $\eps$.

Continuity of $f\mapsto\psi_f$ maps weak convergence of measures to $C^3$ convergence of potentials. The centrist potential bounds therefore persist near $\delta_0$. Let $\mathcal G_\eps$ denote the Gibbs map in Eq.~\eqref{eq:gibbs-map}. Uniform Laplace estimates give
\begin{equation}
 \int y^2\mathcal G_\eps(f)(y)\dd y\leq C\eps
 \label{eq:laplace-bound}
\end{equation}
throughout that neighbourhood. Choose $C$ sufficiently large and $\eps$ sufficiently small. The compact convex set of even measures satisfying $\int y^2f\leq C\eps$ is then invariant under $\mathcal G_\eps$. Schauder's theorem gives a centred fixed point $f^*_{K,r,\eps}$ in this set. Under Assumption~\ref{ass:regular-branch}, the implicit-function theorem continues this fixed point as a $C^1$ branch in $r$. Because $f^*\rightharpoonup\delta_0$ and $\psi_{f^*}\to\psi_0$ in $C^3$, Laplace's method gives
\begin{align}
 f^*_{K,r,\eps}(\sqrt\eps z)\sqrt\eps
 &\longrightarrow
 \sqrt{\frac{\kappa_K(r)}{2\pi}}
 e^{-\kappa_K(r)z^2/2},
 \label{eq:rescaled-gaussian}\\
 \operatorname{Var}(f^*_{K,r,\eps})
 &=\frac{\eps}{\kappa_K(r)}+o(\eps).
 \label{eq:variance-asymptotic}
\end{align}
The convergence is uniform for $r$ in every compact sub-neighbourhood on which Assumption~\ref{ass:centered-well} holds.

To connect this Gibbs profile with the point-platform eigenvalue, we project the marginal seat potential onto translation displacements. Consider a point-platform differentiation $y_i=\delta z_i$ with $z\in\Tspace$. The force on party $i$ is the derivative of its strategic potential at its own location. Moving the evaluation point contributes
\begin{equation}
 \psi_0''(0)z_i=-\kappa_K(r)z_i.
\end{equation}
The change in the party state follows from Eq.~\eqref{eq:Boperator} with the translation distribution $g=-\delta_0'$. For this deformation,
\begin{equation}
 b_g(x)=\partial_yK_r(x,0).
\end{equation}
Differentiating the resulting potential at $y=0$ gives
\begin{equation}
 C_K(r)z_i
 =\frac{z_i}{Q_0(r)}\int\rho(x)\eta_K(s_r(x))
 [\partial_yK_r(x,0)]^2\dd x.
\end{equation}
The point-platform differentiation eigenvalue is therefore
\begin{equation}
 \Lambda_K^{\mathrm{PR}}(r)=C_K(r)-\kappa_K(r).
 \label{eq:Lambda-C-kappa}
\end{equation}
This connects the distributional operator to the point calculation in Appendix~\ref{app:pointlinearisation}.

Finally, we analyse the principal spectral asymptotics of the Rayleigh quotient as $\eps\downarrow0$. Write an odd perturbation as $g=f^*h$ and normalise it by
\begin{equation}
 \int f^*h^2\dd y=1.
\end{equation}
Because $f^*$ is even and $h$ is odd, a Taylor expansion of the kernel at $y=0$ gives
\begin{equation}
 b_g(x)
 =\partial_yK_r(x,0)m_1[g]+R_\eps(x),
 \qquad
 m_1[g]=\int yf^*(y)h(y)\dd y,
 \label{eq:bg-expansion}
\end{equation}
where the even Taylor terms integrate to zero and
\begin{equation}
 \lVert R_\eps\rVert_\infty=o(\sqrt\eps)
\end{equation}
uniformly on the normalised odd unit sphere. The Gaussian moment bounds from Eq.~\eqref{eq:rescaled-gaussian} justify the remainder estimate. In addition,
\begin{equation}
 A_*(x)\to s_r(x),
 \qquad Q_*\to Q_0,
\end{equation}
uniformly in the relevant integrals. The Rayleigh numerator in Eq.~\eqref{eq:nu} is therefore
\begin{equation}
 C_K(r)m_1[g]^2+o(\eps).
 \label{eq:rayleigh-leading}
\end{equation}
Cauchy--Schwarz gives
\begin{equation}
 m_1[g]^2\leq\operatorname{Var}(f^*),
\end{equation}
with equality for the normalised linear mode $h=y/\sqrt{\operatorname{Var}(f^*)}$. Combining Eqs.~\eqref{eq:variance-asymptotic} and~\eqref{eq:rayleigh-leading} gives
\begin{equation}
 \frac{\nu_{K,\eps}(r)}{\eps}
 \longrightarrow\frac{C_K(r)}{\kappa_K(r)}.
 \label{eq:nu-limit}
\end{equation}
Any normalised maximising sequence must approach equality in Cauchy--Schwarz. Its profile component therefore converges to the linear Gaussian mode $h\propto y$, the first Hermite mode. The Gibbs identity then gives
\begin{equation}
 -\frac{(f^*)'}{f^*}=-\frac{(\psi_{f^*})'}{\eps}
 =\frac{\kappa_K(r)y}{\eps}+o(\eps^{-1/2})
\end{equation}
on the concentration scale $y=O(\sqrt\eps)$. After normalisation, the principal density deformation therefore converges to the infinitesimal translation mode $-(f^*)'$. Define
\begin{equation}
 F_K(r,\eps)=\frac{\nu_{K,\eps}(r)}{\eps}-1.
\end{equation}
Equations~\eqref{eq:nu-limit} and~\eqref{eq:Lambda-C-kappa} give
\begin{equation}
 F_K(r,\eps)\longrightarrow
 \frac{\Lambda_K^{\mathrm{PR}}(r)}{\kappa_K(r)}
\end{equation}
uniformly near $r_c(K)$. The transverse point crossing, $\partial_r\Lambda_K^{\mathrm{PR}}(r_c)\neq0$, gives opposite signs at the endpoints of every sufficiently small interval centred on $r_c(K)$. Uniform convergence preserves these endpoint signs for sufficiently small $\eps$. Continuity along the selected branch then gives a zero of $F_K$ inside the interval. On any compact subset of the neighbourhood excluding $r_c(K)$, the limiting function is bounded away from zero, so $F_K$ has no zeros there for sufficiently small $\eps$. Consequently, every choice of a critical ratio in this neighbourhood converges to $r_c(K)$. The odd-sector stability signs away from the limiting threshold follow from Theorem~\ref{thm:distributional-spectral}.

\section{Point-platform eigenvalue, uniqueness, and \texorpdfstring{large-$K$}{large-K} scaling}
\label{app:pointanalysis}

\label{app:pointlinearisation}
The Gaussian point-platform eigenvalue gives a closed-form reference for the finite-width critical boundary. Setting $\sigma_0=1$ temporarily and defining $s(x)=e^{-x^2/(2r^2)}$, the choice probability at the centred state $y_i=0$ satisfies
\begin{equation}
 p_i(x)=\frac{1-(1-s(x))^K}{K},
\end{equation}
which yields the total turnout integral
\begin{equation}
 Q_K(r)=r\sum_{m=1}^K(-1)^{m+1}\binom Km(r^2+m)^{-1/2}.
 \label{eq:QK}
\end{equation}
Reflection symmetry makes all first derivatives of $q_i$ and $Q$ vanish at the centred state. For $j\neq i$, the own-minus-cross curvature of $S_i=q_i/Q$ is therefore
\begin{equation}
 \partial_{ii}S_i-\partial_{ij}S_i
 =\frac1{Q_K}\left[(\partial_{ii}q_i-\partial_{ij}q_i)
 -\frac1K(\partial_{ii}Q-\partial_{ij}Q)\right].
 \label{eq:quotcurv}
\end{equation}
Permutation symmetry identifies this curvature with the differentiation eigenvalue on $\Tspace$. Set $t=r^2$ and define $c_n(z)=1-(1-e^{-z^2/2})^n$. Let $Z_n(t)$ be the positive normalising factor and $M_n(t)$ its normalised second moment:
\[
 Z_n(t)=\int_{\mathbb R}e^{-tz^2/2}c_n(z)\dd z,
 \qquad
 M_n(t)=\frac{\int_{\mathbb R}z^2e^{-tz^2/2}c_n(z)\dd z}{Z_n(t)}.
\]
Differentiating Eq.~\eqref{eq:choice} and evaluating the resulting Gaussian integrals reduces Eq.~\eqref{eq:quotcurv} to
\begin{equation}
 \lambda_{\mathrm{diff}}^{\mathrm{PR}}(K,r)
 =\frac{1}{\sigma_0^2}\Lambda_K^{\mathrm{PR}}(r),
 \qquad
 \Lambda_K^{\mathrm{PR}}(r)
 =\frac{Z_{K-1}(t)}{\sqrt{2\pi}K(K-1)rQ_K(r)}
 \left[(K+t)M_{K-1}(t)-K\right].
 \label{eq:lambdaPR}
\end{equation}
Here adaptation speed is normalised to one. For $K=2$,
\begin{equation}
 \Lambda_2^{\mathrm{PR}}(r)=\frac{\sqrt{r^2+2}}
 {2(r^2+1)(\sqrt{r^2+1}-2\sqrt{r^2+2})}<0.
 \label{eq:k2closed}
\end{equation}
Thus the centred binary point state has a restoring divergence response for every $r>0$. For $K\geq3$, the sign is determined by
\begin{equation}
 G_K(t)=(K+t)M_{K-1}(t)-K.
 \label{eq:GK}
\end{equation}
The unique thresholds include $r_c(3)=0.6630$ and $r_c(5)=0.8476$. For the binary mixed objective,
\begin{equation}
 \Lambda_2^\theta=\theta\Lambda_2^{\rm PR}+(1-\theta)\Lambda_2^{\rm raw},
 \qquad \Lambda_2^{\rm raw}(r)=-\frac{r}{(r^2+1)^{3/2}}+\frac1{2r\sqrt{r^2+2}}.
 \label{eq:theta-point-check}
\end{equation}
The raw-support zero is $r\simeq0.8074$~\citep{LanzettiHajarDorfler2022}. Whenever $\Lambda_2^{\rm raw}>0$, the mixed point state is unstable for $\theta<\Lambda_2^{\rm raw}/(\Lambda_2^{\rm raw}-\Lambda_2^{\rm PR})$.

\label{app:pointmoment}
To prove that the point-platform threshold $r_c(K)$ is unique, observe that every factor in Eq.~\eqref{eq:lambdaPR} outside $[(K+t)M_{K-1}(t)-K]$ is strictly positive. The sign of the point-platform eigenvalue is therefore strictly determined by the sign of $G_K(t)$ in Eq.~\eqref{eq:GK}. To prove that this function has a unique zero, define
\begin{equation}
 H_K(t)=\frac1{M_{K-1}(t)}-1-\frac{t}{K}.
\end{equation}
The critical equation is equivalent to $H_K(t)=0$. Monotone likelihood-ratio ordering gives $M_{K-1}(0)>1$ for $K\geq3$. By contrast, $M_{K-1}(t)\sim t^{-1}$ as $t\to\infty$. Hence $H_K$ changes sign and has at least one root. If $\pi_{n,t}$ is the normalised density proportional to $e^{-tz^2/2}c_n(z)$, differentiation gives
\begin{equation}
 M_n'(t)=-\frac12\operatorname{Var}_{\pi_{n,t}}(Z^2).
\end{equation}
Writing $\kappa_{n,t}=\E Z^4/M_n(t)^2$,
\begin{equation}
 H_K'(t)=\frac{\kappa_{K-1,t}-1}{2}-\frac1K.
\end{equation}
The density $\pi_{n,t}$ is symmetric and unimodal. Its kurtosis is at least $9/5$~\citep{KlaassenMokveldVanEs2000}. Therefore,
\begin{equation}
 H_K'(t)\geq\frac25-\frac1K>0
 \qquad(K\geq3),
\end{equation}
so $H_K$ is strictly increasing. Together with the sign change, this proves existence and uniqueness of $r_c(K)$.

We next establish the large-$K$ asymptotic expansion of the critical boundary. Let $Z_t\sim N(0,t^{-1})$ and define
\begin{align}
 R_n(t)&=\E\left(1-e^{-Z_t^2/2}\right)^n,\\
 L_n(t)&=\E\left[Z_t^2\left(1-e^{-Z_t^2/2}\right)^n\right],\\
 \Delta_n(t)&=L_n(t)-R_n(t)/t.
\end{align}
At the critical point, set $n=K-1$. The critical equation becomes
\begin{equation}
 \frac{K(1-t)+t}{t(K+t)}
 =\frac{\Delta_{K-1}(t)}{1-R_{K-1}(t)}.
 \label{eq:criticaltail}
\end{equation}
Applying the substitutions $y=e^{-z^2/2}$ and then $y=u/n$ gives the following uniform expansions for $t=1+o(1)$:
\begin{align}
 R_n(t)
 &=\sqrt{\frac{t}{\pi}}\Gamma(t)n^{-t}(\log n)^{-1/2}
 \left[1+\frac{\psi(t)}{2\log n}+O((\log n)^{-2})\right],\\
 \Delta_n(t)
 &=\sqrt{\frac{t}{\pi}}\Gamma(t)n^{-t}
 \left[2\sqrt{\log n}
 -\frac{\psi(t)+t^{-1}}{\sqrt{\log n}}
 +O((\log n)^{-3/2})\right],
 \label{eq:tailasym}
\end{align}
where $\psi$ is the digamma function. Equation~\eqref{eq:criticaltail} first implies $t_c\to1$. Substitute Eq.~\eqref{eq:tailasym} and use $\psi(1)=-\gamma_E$. Iterating the resulting approximation once gives
\begin{align}
 1-r_c(K)^2
 =&\frac{2}{\sqrt\pi}\frac{\sqrt{\log K}}{K}
 -\frac1K
 +\frac{\gamma_E-1}{\sqrt\pi K\sqrt{\log K}}
 \nonumber\\
 &+O\!\left(\frac{1}{K(\log K)^{3/2}}
 +\frac{(\log K)^2}{K^2}\right).
 \label{eq:largeKexpansion}
\end{align}
Dividing by $1+r_c(K)$ gives
\begin{equation}
 r_c(K)\to1,\qquad 1-r_c(K)\sim\frac{\sqrt{\log K}}{\sqrt\pi K}.
 \label{eq:largeKleading}
\end{equation}

\begin{figure}[htp]
\centering
\includegraphics[width=0.70\textwidth]{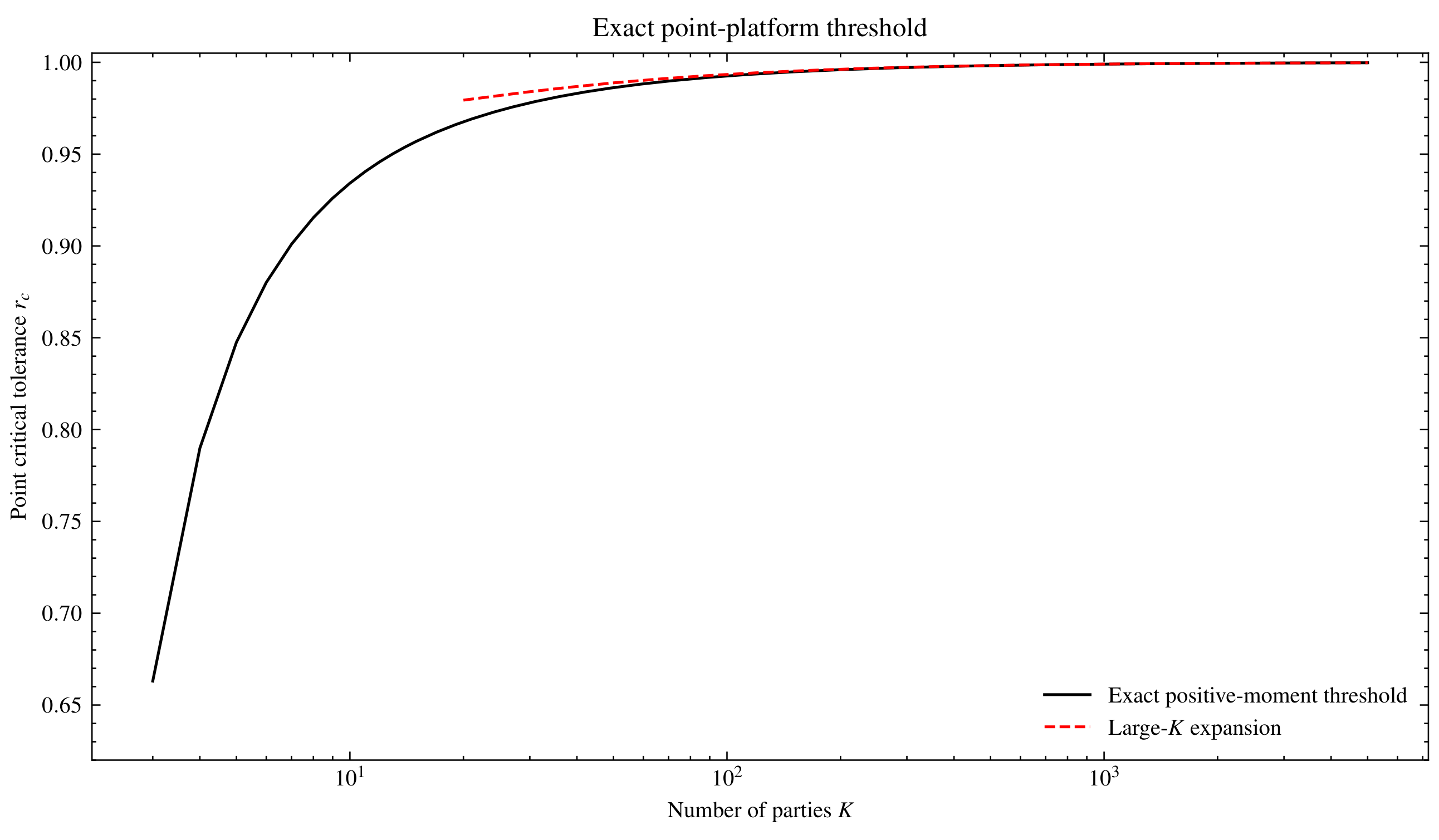}
\caption{Exact point-platform tolerance threshold. The solid curve gives the unique root $r_c(K)$ of the positive-moment equation. The dashed curve gives the large-$K$ expansion. At the symmetric reference state, the unstable range expands with the number of equally viable parties, while $r_c(K)$ approaches one.}
\label{fig:critical}
\end{figure}

\section{Institutional gain and geographic curvature}
\label{app:institution}

\label{app:institutionproof}
Theorem~\ref{thm:effective-temperature} follows from the action of the seat-map Jacobian on party-divergence directions. Party anonymity means that relabelling vote shares relabels seat shares:
\begin{equation}
 \mathcal R(Pv)=P\mathcal R(v).
\end{equation}
Differentiating at the symmetric vector $v^*=K^{-1}\mathbf{1}$, where $\mathbf{1}=(1,\ldots,1)^\top$, gives $D\mathcal R(v^*)P=PD\mathcal R(v^*)$ for every permutation. A matrix that commutes with every permutation has the form
\begin{equation}
 D\mathcal R(v^*)=aI+b\1\1^\top.
\end{equation}
The rank-one term vanishes on $\Tspace$. Hence the Jacobian restricts to $g_RI_{\Tspace}$, where $g_R=a$. Theorem~\ref{thm:effective-temperature} assumes the local responsiveness condition $g_R>0$.

At a homogeneous diagonal state, every mass-preserving first-order change in valid-vote shares lies in $\Tspace$. The marginal seat potential under $R$ is therefore $g_R$ times its PR counterpart. An additive constant may remain, but it does not affect the Wasserstein drift. The zero-current condition is
\begin{equation}
 f^*\partial_y(g_R\psi_{\mathrm{PR}})-\eps\partial_yf^*=0,
\end{equation}
which is the PR Gibbs equation at effective noise parameter $\theta_R=\eps/g_R$ (an effective ideological temperature). Consider a reflection-odd perturbation in the divergence subspace. Its national first-order vote response vanishes. The relevant integrand is the odd function $b_g(x)$ multiplied by even diagonal coefficients. Hence $D^2\mathcal R[Dv,Dv]$ vanishes at the homogeneous reference state. The remaining strategic linearisation is multiplied by $g_R$, while diffusion is unchanged. On the corresponding diagonal state,
\begin{equation}
 \Lop_R^{\mathrm{odd}}(\eps)
 =g_R\Lop_{\mathrm{PR}}^{\mathrm{odd}}(\eps/g_R),
\end{equation}
which proves Theorem~\ref{thm:effective-temperature}.

Under smooth plurality, the softmax Jacobian is $(\beta/K)I$ on $\Tspace$, so $g_{\mathrm{FPTP}}=\beta/K$. For the MMP surrogate, define
\begin{equation}
 \lambda_*=\left[1+e^{-\gamma(\alpha-1)/K}\right]^{-1},
 \qquad
 u_*=\frac1K+\frac1\gamma\log\!\left(1+e^{\gamma(\alpha-1)/K}\right).
\end{equation}
At symmetry, a first-order perturbation in the divergence subspace changes the unnormalised entitlement according to
\begin{equation}
 \delta u_i=(1-\lambda_*)\delta P_i+\lambda_*\alpha\delta C_i,
\end{equation}
where $\delta C_i=(\beta/K)\delta v_i$. Normalising by $\sum_i u_i=Ku_*$ gives
\begin{equation}
 g_{\mathrm{MMP}}
 =\frac{(1-\lambda_*)+\lambda_*\alpha\beta/K}{Ku_*}>0.
 \label{eq:mmpgain}
\end{equation}

\begin{proposition}[Geographic activation]
\label{prop:geography}
Suppose districts occur in equal-weight reflected pairs, $\rho_{\bar d}(x)=\rho_d(-x)$, treated identically by the rule. The centred point configuration is an equilibrium. For distributed parties, consider an even diagonal stationary state for the specified densities and rule. An odd deformation induces opposite first-order vote responses in paired districts; their quadratic products preserve sign. A nonzero seat-map Hessian contraction along these response directions enters the local divergence operator.
\end{proposition}

For point platforms $\bm y$, write $\widehat S_i^R(\bm y)=\mathcal S_i^R(\bm v(\bm y))$. The dependence proceeds from $a_j(x)=K_\sigma(x,y_j)$ through district support and turnout to $v_{\ell d}(\bm y)=q_{\ell d}(\bm y)/T_d(\bm y)$. The first derivative, holding other platforms fixed, is
\[
 \partial_{y_a}\widehat S_i^R=\sum_{\ell,d}
 \frac{\partial\mathcal S_i^R}{\partial v_{\ell d}}\frac{\partial v_{\ell d}}{\partial y_a}.
\]
Differentiating both factors with respect to $y_b$ gives
\begin{align}
 \frac{\partial^2\widehat S_i^R}{\partial y_a\partial y_b}
 =&\sum_{\ell,d}\frac{\partial\mathcal S_i^R}{\partial v_{\ell d}}
 \frac{\partial^2v_{\ell d}}{\partial y_a\partial y_b}\nonumber\\
 &+\sum_{\ell,d,m,e}\frac{\partial^2\mathcal S_i^R}{\partial v_{\ell d}\partial v_{me}}
 \frac{\partial v_{\ell d}}{\partial y_a}\frac{\partial v_{me}}{\partial y_b}.
 \label{eq:curvaturedecomp}
\end{align}
Here $a,b$ index the moving parties. The first line is behavioural vote curvature; the second is the seat-map Hessian applied to local vote responses, evaluated at $\bm v(\bm y)$. At an even diagonal stationary density, let $h_i$ perturb party $i$ alone and $g_j$ perturb party $j$ alone. The same chain rule becomes
\begin{equation}
 D^2(S_i^R\circ v)[h_i,g_j]
 =D S_i^R[v]D^2v[h_i,g_j]+D^2S_i^R[v][Dv[h_i],Dv[g_j]].
 \label{eq:functional-curvature}
\end{equation}
If districts form reflected pairs, $\rho_{\bar d}(x)=\rho_d(-x)$, and $g$ is odd, then
\begin{equation}
 Dv_{\bar d}[g]=-Dv_d[g].
\end{equation}
Equal district weights and identical treatment make the first derivative cancel within each reflected pair. This gives zero force on centred point parties. For densities, reflection preserves evenness, and the linearisation is evaluated at the stationary profile specified in the proposition. When a district density is shifted, $Dv_d[g]$ is nonzero except in degenerate response configurations. Both odd response factors in Eq.~\eqref{eq:functional-curvature} reverse sign under reflection, so their product is unchanged. Summing the mixed derivatives over party perturbations with coefficients $z_j$, where $\sum_jz_j=0$, gives the response in the divergence subspace. A nonzero institutional Hessian contraction in this response enters the local operator.

\paragraph{Binary cancellation of institutional curvature.}
\label{app:binary-curvature}
For $K=2$, write the district vote vectors as $v_d=(1/2+u_d,1/2-u_d)$ and define $\widehat S_i^R(u)=S_i^R(v(u))$. At a diagonal party configuration, $u=0$ in every district, regardless of its voter density. Exchanging the two party labels sends $u$ to $-u$. Party anonymity and $\widehat S_1^R+\widehat S_2^R=1$ therefore imply
\begin{equation}
\begin{aligned}
 \widehat S_1^R(-u)&=1-\widehat S_1^R(u),\\
 D^2\widehat S_1^R(0)&=0.
\end{aligned}
\label{eq:binary-seat-curvature}
\end{equation}
The second identity follows by differentiating the first twice at zero. It holds for any smooth, normalised, party-anonymous binary seat map, including the three numerical rules. The institutional Hessian contraction in Eq.~\eqref{eq:curvaturedecomp} therefore vanishes at the binary tie, including with reflected heterogeneous districts. The behavioural term $DS_i^R\,D^2v$ remains available and must be evaluated to determine stability. The identity is local to the tie; nonlinear institutional responses can differ away from it. The point-platform restoring rates below use the full strategic force.

\section{Numerical methods and reproducibility}
\label{app:numerics}

This appendix specifies the mixed-objective stationary, spectral and nonlinear calculations in Appendix~\ref{app:theta-numerics}, followed by the seat-share maximisation and geographic benchmarks. The simulators differentiate $(1-\theta)\bar q_i+\theta\mathcal S_i^R$, retaining seat shares as separate observables.

All numerical experiments use synthetic electorates and party states. The archive contains the exact choice operator, electoral maps, point-platform and particle dynamics, diagonal-state solver, and spectral solver. It also records the software environment, random seeds, raw outputs, and figure-generation commands.

\subsection{Mixed-objective stationary states, spectra and dynamics}
\label{app:theta-numerics}

For the mixed-objective experiments in Section~\ref{sec:theta-results}, voters follow a standard Gaussian distribution, integrated over $\R$ by 400-node Gauss--Hermite quadrature, except for the finite-particle runs described below. Party densities use 401 equally spaced nodes $y_j$ on $\Omega=[-2.2,2.2]$, with trapezoidal weights $w_j$ and Gaussian satisfaction kernel $\exp[-(x-y)^2/(2r^2)]$. At each parameter combination, damped iteration of Eq.~\eqref{eq:gibbs-map}, with damping factor $0.12$, computes an even common profile. Reflection symmetry and unit mass are imposed during this stationary iteration. The stopping criterion is $\sum_jw_j|\mathcal G_{\eps,\theta}(f)_j-f_j|<10^{-10}$, tightened to $10^{-12}$ for the reference states used in nonlinear runs. Parameter scans reuse nearby converged profiles as initial guesses. This procedure selects a stationary branch; it does not establish uniqueness of the fixed point.

The strategic eigenproblem in Eq.~\eqref{eq:theta-nu} uses coordinates $v_j=\sqrt{w_j/f^*_\theta(y_j)}\,g_j$, so its weighted denominator becomes the Euclidean norm. Symmetric Lanczos iteration, restricted to reflection-odd vectors, computes the largest eigenvalue with tolerance $10^{-11}$. The binary map samples 37 equally spaced values of $r\in[0.2,1.1]$ and 21 values of $\theta\in[0,1]$ at $\eps=0.002$. For each $K\in\{2,3,4,5\}$ and sampled $\theta$, sign changes of $\nu/\eps-1$ bracket critical tolerances; the multiparty search uses 13 initial points on $[0.5,1.1]$. Brent's method refines the brackets with absolute tolerance $2\times10^{-6}$. A missing crossing denotes only the absence of a detected root in the stated window. At $K=2$ and $r=0.6$, a separate search in $\theta$ uses tolerance $10^{-11}$ to resolve the nearly neutral dynamics at $\theta_c$.

Temporal growth rates retain the mobility factor in Eq.~\eqref{eq:MHfactor}. We discretise the no-flux drift--diffusion operator using the same conservative scheme as the nonlinear evolution below and restrict it to odd parity. In the weighted coordinates above, let $L_0$ be the symmetric generator with the stationary potential held fixed, and let $C$ represent the strategic operator. The full linearised generator is $L_0(I-C/\eps)$. Factoring $-L_0=U^\top U$ reduces its spectrum to that of the symmetric matrix $U(C/\eps-I)U^\top$; its largest eigenvalue gives $\lambda_{\rm odd}$. Across the 777 binary map points, the archived signs of $\lambda_{\rm odd}$ agree with those of $\nu/\eps-1$. The largest recorded stationary and transformed generator eigenpair residuals are $2.22\times10^{-11}$ and $1.54\times10^{-12}$, respectively.

For the full binary dynamics, set $K=2$, $r=0.6$ and $\eps=0.002$. Each objective weight has its own common state $f^*_\theta$. With $s_\theta^2=\int_\Omega y^2f^*_\theta(y)\dd y$, the initial profiles are
\begin{equation}
 f_i(y,0)=f^*_\theta(y)
 \left[1+\delta z_i\tanh\!\left(\frac{y}{s_\theta}\right)\right],
 \qquad \delta=0.02,\qquad (z_1,z_2)=(-1,1).
 \label{eq:theta-initial}
\end{equation}
The integral of the perturbation vanishes because $f^*_\theta$ is even and $\tanh(y/s_\theta)$ is odd. The bracket lies between $1-\delta$ and $1+\delta$, so both profiles remain strictly positive. The same statements hold for the reflected grid and symmetric quadrature weights. We evolve the five cases $\theta\in\{1,\theta_c+0.04,\theta_c,\theta_c-0.04,0\}$ without subsequent symmetry projection.

At each time step, the exact choice probabilities and their unilateral derivatives determine both parties' mixed potentials simultaneously. Writing $\Psi_{i,j}^n=\psi_{i,\theta}[f^n](y_j)$, $a_{i,j+1/2}^n=(\Psi_{i,j+1}^n-\Psi_{i,j}^n)/\eps$, and $b(a)=a/(e^a-1)$ with $b(0)=1$, the Scharfetter--Gummel face flux is
\begin{equation}
 J_{i,j+1/2}^{n+1}
 =\frac{\eps}{\Delta y}
 \left[b(-a_{i,j+1/2}^n)f_{i,j}^{n+1}
       -b(a_{i,j+1/2}^n)f_{i,j+1}^{n+1}\right].
 \label{eq:theta-sg-flux}
\end{equation}
We set the outer boundary fluxes to zero and solve
$w_j(f_{i,j}^{n+1}-f_{i,j}^{n})/\Delta t=J_{i,j-1/2}^{n+1}-J_{i,j+1/2}^{n+1}$
with $\Delta t=0.1$. Each tridiagonal solve is implicit in density and holds the current potential fixed. The update preserves mass and positivity without clipping or renormalising the evolving profiles. A noncritical run stops after $t\geq50$ when the maximum partywise weighted $L^1$ norm of the full discrete right-hand side falls below $10^{-9}$, or at $t=1600$. The critical run always reaches the full horizon and is not classified as a converged endpoint. We separately record the $L^1$ discrepancy from each party's Gibbs density evaluated at the current coupled potentials.

Growth and decay are measured by ordinary least squares of $\log P(t)$ on time over $5\leq t\leq35$, retaining observations with $10^{-12}<P(t)<0.05$ and cases with $|\lambda_{\rm odd}|>10^{-7}$. For the two unstable endpoints, we repeat the evolution with $\Delta t=0.05$ and restart from terminal densities multiplied by $e^{0.03y}$ and $e^{0.01y}$ for parties 1 and 2, respectively, normalising each tilted initial profile once before integration. These checks recover terminal spread and turnout to within $2\times10^{-10}$.

Resolution checks use $(N_y,N_x,L)=(301,240,2.2)$, $(401,400,2.2)$ and $(601,600,2.6)$ for the critical tolerances at $(K,\theta)=(2,0),(2,0.5),(3,1),(5,1)$. At these four baseline-grid roots, stationary iterations started from Gaussian widths $0.06$ and $0.5$ converge to profiles differing by less than $1.4\times10^{-14}$ in weighted $L^1$. Growth-rate checks at $r=0.6$ use 201, 401 and 801 ideology nodes with 400 voter quadrature nodes for $(K,\theta)=(2,0),(2,0.28),(2,0.36),(2,1),(5,0),(5,1)$. Additional critical-tolerance calculations at $\eps=0.001,0.002,0.004$ assess diffusion sensitivity for the four root cases. The archived scripts and outputs retain these case definitions and the individual residuals.

Figure~\ref{fig:theta-dynamics}(c) uses a separate finite-particle approximation to the reflected diffusion in Appendix~\ref{app:wellposed}, at $K=2$, $r=0.6$, $\eps=0.002$ and $\theta=\theta_c-0.04=0.283638537\ldots$. Each party has $N=200$ particles. We recompute the even common profile on the same 401-node ideology grid using 160-node voter quadrature and stationary tolerance $10^{-12}$. Linear interpolation of its discrete cumulative mass at quantiles $(a-1/2)/N$ gives positions $q_a$; both parties start at $X_{i,a}^0=(q_a-q_{N+1-a})/2$. Thus the particle initial states coincide and are exactly reflection-symmetric, without the odd perturbation in Eq.~\eqref{eq:theta-initial}. At every step, satisfaction is averaged directly over each party's empirical measure $\mu_i^{N,n}=N^{-1}\sum_a\delta_{X_{i,a}^n}$, and exact choice derivatives give the drift $b_i=\partial_y\psi_i^{R,\theta}$ with $R$ equal to threshold-free PR. The reflected Euler--Maruyama update is
\begin{equation}
 X_{i,a}^{n+1}=\mathcal R_\Omega\!\left(
 X_{i,a}^n+\Delta t\,b_i(X_{i,a}^n;\bm\mu^{N,n})
 +\sqrt{2\eps\Delta t}\,\xi_{i,a}^n\right),
 \label{eq:theta-particle-step}
\end{equation}
where the $\xi_{i,a}^n$ are independent standard normal draws and $\mathcal R_\Omega$ folds boundary overshoots back into $[-2.2,2.2]$ by repeated reflection. We use $\Delta t=0.1$, run to $t=800$ without early stopping, and record every ten steps. NumPy's default random generator uses seeds 0--11.

The projected coordinates are $d(t)=(m_2(t)-m_1(t))/2$, with $m_i=N^{-1}\sum_aX_{i,a}$, and turnout evaluated from the current empirical measures. Seven archived runs end with $d>0$ and five with $d<0$; seeds 0 and 1 highlight one trajectory of each sign. The dashed curves plot $(P(t),T(t))$ from the deterministic $\theta_c-0.04$ run and its party-exchanged image $(-P(t),T(t))$. For these reflected binary density profiles, $P=|d|$. These curves are reference trajectories, with stars at the computed stationary endpoints. The finite-particle runs illustrate noise-induced departure from the common state; twelve seeds do not estimate branch-selection probabilities or establish convergence in particle number or time step.

\subsection{Seat-share benchmarks and geographic experiments}

For the numerical PR and MMP maps, the proportional entitlement is
\[
 P_i=\frac{\bar v_i\,s_k(\bar v_i-\tau)}{\sum_j\bar v_j\,s_k(\bar v_j-\tau)},
 \qquad s_k(u)=(1+e^{-ku})^{-1},\qquad \tau=0.05,\ k=80.
\]
Threshold-free PR uses $P_i=\bar v_i$. The plurality map is independent of this gate.

In the distributional spectral calculations, the benchmark electorate is standard Gaussian and is integrated by 160-node Gauss--Hermite quadrature. Party ideology uses 801 evenly spaced points on $[-2.2,2.2]$. Damped fixed-point iteration solves Eq.~\eqref{eq:gibbs-map} until its $L^1$ residual falls below $10^{-10}$. A similarity transform uses the weighted denominator $\int g^2/f^*$. The operator in Eq.~\eqref{eq:Boperator} is then applied without forming a dense matrix. Symmetric Lanczos iteration computes its largest odd eigenvalue. Brent's method locates zeros of $\nu/\eps-1$. The manuscript phase diagram uses
\begin{equation}
 K\in\{3,4,5\},
 \qquad
 \eps\in\{5\!\times\!10^{-4},10^{-3},2\!\times\!10^{-3},4\!\times\!10^{-3},8\!\times\!10^{-3}\}.
\end{equation}
Every reported root is bracketed by values with opposite signs of $\nu/\eps-1$. A conservative finite-volume discretisation of Eq.~\eqref{eq:odd-linearized-pde} also gives growth below the root and decay above it. We repeat representative cases with 401--1001 ideology nodes and 100--200 Gaussian nodes. We also vary the domain from $[-1.8,1.8]$ to $[-2.4,2.4]$. These checks leave the reported ratios unchanged at the displayed precision.

For the dynamic particle and district experiments, the geographic benchmark uses $K=5$ parties and $D=9$ districts, with satisfaction width $\sigma=0.65$. District means occupy normal-quantile positions scaled by $g$, with within-district standard deviation $\sqrt{1-g^2}$ over the reported geography range. Each district density is truncated and normalised on $[-4,4]$. Reflection pairing keeps the national density centred; its standard deviation and shape vary with $g$. On the 61-node voter grid, the national standard deviation is $0.9995$ at $g=0$, $0.9795$ at $g=0.55$, $0.9668$ at $g=0.70$, and $0.9507$ at $g=0.85$. Point-platform positions are drawn independently from $N(0,0.12^2)$ and sorted by party label, with five initialisations for each parameter value. Explicit-gradient integration uses time step $0.15$ and maximum horizon $42$. The numerical PR and MMP comparisons apply a smooth five-percent electoral threshold. The analytical PR thresholds use exact threshold-free PR. Smooth plurality uses $\beta=24$. The stylised MMP map uses $\alpha=0.6$ and $\gamma=45$. The order parameter is the seat-weighted standard deviation of party ideological means:
\begin{equation}
 P=\left[\sum_iS_i(m_i-\bar m_S)^2\right]^{1/2},
 \qquad
 \bar m_S=\sum_iS_im_i.
\end{equation}
The archive also reports abstention and effective electoral and parliamentary party numbers. Finite-particle convergence, time-step checks, initialisation checks, MMP sensitivity and hard-winner limits assess the robustness of the dynamic benchmarks. The analytical phase boundaries are determined separately by the odd-sector spectral criterion.

\subsection{Geographic comparisons at matched national density}
\label{app:matched-geography}

For $g\in\{0,0.55,0.70,0.85\}$, define the national density $\bar\rho_g=\sum_dw_d\rho_{d,g}$. Each geographic configuration is paired with a homogeneous control having $\rho^{\mathrm{hom}}_{d,g}=\bar\rho_g$ in every district. The national densities agree to within $1.2\times10^{-16}$ on the numerical grid. Both members use the same rule, parameters, and initial positions for seeds 0--4, and both run to $T=42$ without early stopping. Thus the paired difference $P_g-P_g^{\mathrm{hom}}$ measures the effect of district organisation conditional on that national density. Changes in the homogeneous controls across $g$ record the effect of the changing national distribution along this parameter family.

At $g=0$, the paired outcomes coincide. At $g=0.55$, $0.70$, and $0.85$, the mean paired differences are respectively $0.258$, $0.289$, and $0.308$ for plurality; $0.058$, $0.069$, and $0.092$ for MMP; and $-0.016$, $-0.008$, and $0.030$ for PR. At $g=0.70$, the plurality--PR gap is $0.319$ with districts and $0.022$ in the homogeneous control. The larger separation between these rules therefore persists after matching the national electorate within each pair. Figure~\ref{fig:matched-geography} reports the comparisons and the variation of the paired differences across initialisations.

\subsection{Binary geographic comparison}
\label{app:binary-geography}

\begin{table}[htp]
\centering
\caption{Binary point-platform divergence rates at the centre in the heterogeneous-district configurations; $D=9$ and $\sigma=0.65$. The $g=0$ row is homogeneous. All entries are negative.}
\label{tab:binary-local-rates}
\begin{tabular}{rrrr}
\toprule
$g$ & PR & Plurality & MMP\\
\midrule
0.00 & $-0.284908$ & $-3.418901$ & $-0.285125$\\
0.55 & $-0.350048$ & $-4.200582$ & $-0.350314$\\
0.70 & $-0.401571$ & $-4.818856$ & $-0.401876$\\
0.85 & $-0.480145$ & $-5.761743$ & $-0.480510$\\
\bottomrule
\end{tabular}
\end{table}

The binary comparison uses $K=2$, $D=9$, $\sigma=0.65$, the same three seat maps, $g\in\{0,0.55,0.70,0.85\}$, and seeds 0--4. Its 60 matched pairs use identical initial positions within each pair and run to $T=42$ with time step $0.15$ and no early stopping. Alongside $P$, we record the distance $d(t)=|y_1(t)-y_2(t)|$ and the final maximum absolute strategic gradient. The maximum $P(42)$ is $9.41\times10^{-7}$. At $g=0.70$ and $0.85$, halving the time step at fixed $T=42$ gives 60 additional runs; the largest absolute changes in $P$ and $d$ are $1.22\times10^{-7}$ and $2.44\times10^{-7}$, respectively. In 54 baseline runs, the final gradient exceeds $10^{-8}$. Extending every affected pair gives 80 runs at $T=84$, with maximum $P=7.97\times10^{-12}$ and maximum gradient $3.82\times10^{-12}$. The longer runs resolve the remaining finite-time separation as continued convergence towards coincident platforms.

For a local diagnostic, define the own-party strategic force and its divergence rate by
\begin{equation}
\begin{aligned}
 G_i(y)&=\partial_{y_i}S_i^R(v(y)),\\
 \lambda_{\mathrm{div}}&=e^\top DG(0)e,
 \qquad e=(1,-1)^\top/\sqrt{2}.
\end{aligned}
\label{eq:binary-local-rate}
\end{equation}
Reflection pairing makes $G(0)=0$, and party anonymity makes $e$ an eigenvector of $DG(0)$. Automatic differentiation gives the rates in Table~\ref{tab:binary-local-rates}. Both the divergence and common-shift eigenvalues are negative for all district configurations and matched homogeneous controls. At $g=0.70$, the homogeneous-control divergence rates are $-0.283994$, $-3.407922$ and $-0.284209$ for PR, plurality and MMP, respectively. Comparing them with the table isolates a faster local restoring response due to district organisation at this national density. Central differences of the force with displacement $10^{-5}$ agree with the automatic derivatives within $8.3\times10^{-8}$. Increasing the voter grid from 61 to 121 nodes at $g=0.70,0.85$ changes the rates by less than $7.7\times10^{-7}$. The Euler multipliers $1+0.15\lambda_{\mathrm{div}}$ lie strictly between zero and one in all baseline cases.

\end{document}